\documentclass[iop,apj]{emulateapj}
\usepackage{graphicx}
\usepackage{epsf}
\usepackage{natbib}
\usepackage{amsmath}
\usepackage{color}
\usepackage{times}
\usepackage{microtype}
\usepackage{float}
\usepackage[T1]{fontenc}
\usepackage{pslatex}

\newcommand \gta {\mathrel{\vcenter
     {\hbox{$>$}\nointerlineskip\hbox{$\sim$}}}}
\newcommand \ms {M$_\odot$}

\newcommand\kms{km~s$^{-1}$}
\newcommand\gcm{g~cm$^{-3}$}

\def\gtsim{\lower.5ex\hbox{$\buildrel > \over\sim$}}
\def\ltsim{\lower.5ex\hbox{$\buildrel < \over\sim$}}

\def\apj{Ap. J.}
\def\apjl{Ap. J. Lett.}
\def\apjs{Ap. J. Supp.}

\def\nat{Nature}
\def\mnras{Mon. Not. Royal. Soc.}
\def\aap{Astron. \& Astroph.}

\begin{document}

\title{The Role of the Magnetorotational Instability in Massive Stars}

\author{ J. Craig Wheeler\altaffilmark{1}, Daniel Kagan\altaffilmark{2,3}, Emmanouil Chatzopoulos\altaffilmark{4,5}}
\email{wheel@astro.as.utexas.edu}
\altaffiltext{1}{Department of Astronomy, University of Texas at Austin, Austin, TX, USA}
\altaffiltext{2}{Racah Institute of Physics, Hebrew University of Jerusalem, Jerusalem 91904, Israel}
\altaffiltext{3}{Raymond and Beverly Sackler School of Physics and Astronomy, 
Tel Aviv University, Tel Aviv 69978, Israel}
\altaffiltext{4}{Department of Astronomy \& Astrophysics and FLASH Center for Computational Science,
University of Chicago, Chicago, IL, 60637, USA}
\altaffiltext{5}{Enrico Fermi Fellow}

\begin{abstract}

The magnetorotational instability (MRI) is key physics in accretion disks and is widely 
considered to play some role in massive--star core collapse. Models of rotating massive 
stars naturally develop very strong shear at composition boundaries, a necessary condition 
for MRI instability, and the MRI is subject to triply--diffusive destabilizing effects in 
radiative regions. We have used the MESA stellar evolution code to compute magnetic effects 
due to the Spruit--Taylor mechanism and the MRI, separately and together, in a sample of 
massive star models. We find that the MRI can be active in the later stages of massive star 
evolution, leading to mixing effects that are not captured in models that neglect the MRI. 
The MRI and related magneto--rotational effects can move models of given ZAMS mass 
across ``boundaries'' from degenerate CO cores to degenerate O/Ne/Mg cores and from 
degenerate O/Ne/Mg cores to iron cores, thus affecting the final evolution and the physics 
of core collapse. The MRI acting alone can slow the rotation of the inner core in general 
agreement with the observed ``initial" rotation rates of pulsars. The MRI analysis suggests 
that localized fields $\sim 10^{12}$ G may exist at the boundary of the iron core. With 
both the ST and MRI mechanisms active in the 20 \ms\ model, we find that the helium 
shell mixes entirely out into the envelope. Enhanced mixing could yield a population of 
yellow or even blue supergiant supernova progenitors that would not be standard SN~IIP.

\end{abstract}


\section{Introduction}
\label{intro}

One of the major unsolved problems of stellar evolution
is the effect of differential rotation on the magnetic field 
structure of stars and the feedback of that magnetic field on 
the stellar structure and evolution. The role of rotation in stars is 
well-studied if not fully understood \citep{VonZeipel, GS67, Fricke69, 
Tassoul78, Endal81, M09,MM14}.
Some of that work includes the effects of magnetic fields 
\citep[and references therein]{M09}, but this remains a major 
challenge requiring fully three-dimensional studies. Even the
status of the solar rotation and magnetic field remains a major 
issue \citep{CD96,howe09}. The late stages of stellar evolution where 
direct relevant observations are scarce is even more of a challenge.
The actual amount of angular momentum and magnetic field of the 
iron core has obvious implications for the creation of new--born 
neutron stars and for black holes.

Spruit (1999, 2002) presented various magnetic instabilities
that could be involved in stellar evolution and prescriptions 
for treating them, including the Tayler instability \citep{Tayler1973} 
and the magneto--rotational instability \citep[MRI;][]{Vel59, 
chandra60, acheson78, BH91, BH98}. Spruit emphasized the 
nature and role of the Tayler instability in which a toroidal 
field could be perturbed, twisted, and sheared to produce a 
radial field. This is a pinch-type instability of a toroidal magnetic 
field in differentially rotating stellar radiative zones that is 
predicted to result in large-scale fluid motion in the star. Spruit 
gave a prescription for the equilibrium field structure, in particular 
the ratio of the radial and toroidal fields, and for the magnetic 
viscosity that would result from the drag associated with the 
radial component of the field interacting with shear in the star. 
We refer to these effects collectively as the Spruit--Tayler (ST) mechanism.
\citet{HWS05} \citep[see also][]{MM04,Pe05,Ca07,Suijs08,paxton11,
paxton13,Brott11,Ekstrom12,CW12,YDL12} incorporated the 
ST magnetic viscosity prescription in a one--dimensional 
stellar evolution code that had previously been used to explore the 
effect on angular momentum transport of a wide variety of classical 
fluid instabilities \citep{HLW00}. \citet{HWS05} concluded that the 
ST magnetic viscosity would tend to damp the rotation rate of the iron 
core that formed in the final stages of evolution of massive stars 
by a factor of 30 - 50 compared to computations that did not account 
for magnetic torques and that more massive stars would have more 
rapidly rotating iron cores. A variety of issues concerning the ST
mechanism remain open. We return to that topic in \S \ref{discuss}.

Although the MRI has been thoroughly explored in the context of accretion 
disks, it also applies to quasi-spherical objects, e.g. stars \citep{BH94}. 
The MRI is widely considered to play some role in core collapse \citep{aki03, 
MST06, MST07, Ober09,SY14}, but its role in stellar evolution has been substantially 
neglected. This is in part because a threshold shear is required to trigger the MRI 
and the MRI tends to be stabilized by strong thermal and composition gradients 
\citep{M09}. On the other hand, models of rotating massive stars naturally develop 
strong shear at composition boundaries, and the MRI is subject to 
triply-diffusive destabilizing effects in radiative regions \citep{acheson78,MBS04}. 
The MRI grows exponentially rapidly when unstable and can be active in 
the Sun \citep{PM07,M11,KW14}. We argue here that the MRI should 
also be considered in the context of the evolution of massive stars. 

\citet{HWS05} neglect the MRI, but the resulting models tend to give 
very strong radial gradients in the angular velocity in the final 
stages of the evolution (see Figures 2 and 3 in Heger et al.
for the corresponding specific angular momentum gradient distributions). 
These sharp gradients arise at the composition boundaries of the 
``onion-skin" layers that are also the boundaries between (possibly 
extinct) convective cores and outer radiative layers that may once 
themselves have been involved in convective burning. These sharp 
boundaries are stabilized against Kelvin--Helmholtz instabilities by 
the associated composition gradients, but they may be unstable to 
interface dynamos \citep{brun05} or the MRI. One question is whether 
or not such sharp gradients in angular velocity would have developed
in the first place had the MRI been considered as the star evolved 
on and after the main sequence.   

Where in the geometry various instabilities occur is a 
major issue. As noted by \citet{spruit99}, the Tayler instability disappears 
on the equator and shows its most characteristic behavior near the rotation
axis. The MRI may be most active near the equator in radiative shearing regions
where the shear is strong and weaker at the poles, but in the tachocline
and convective envelope of the Sun the MRI tends to be supressed at low
latitudes \citep{PM07,M11,KW14}. In the following, we will neglect these considerations
due to the restrictions of a spherically--symmetric evolution code, but return
to them in \S \ref{discuss}. 

In \S \ref{properties} we present the instability criterion for the MRI, the resulting 
expressions for viscosity and diffusion coefficients that transport angular momentum 
and mix compositions, and our treatment of the growth and saturation of the 
magnetic field. Section \ref{results} describes our use of the MESA code and gives our 
results, and \S \ref{discuss} presents a discussion and conclusions.

\section{Physical Properties of the ST and MRI Mechanisms}
\label{properties}

We first define a number of terms that will be employed in the subsequent discussion.
The angular velocity is $\Omega$ and $q = d \ln \Omega/d \ln r$ is the radial shear.
The Alfv{\'e}n frequency is $\omega_A$. Assuming the toroidal field to dominate, the 
Alfv{\'e}n frequency and Alfv{\'e}n velocity are:
\begin{equation}
\label{alfven}
\omega_{ \rm A} = \frac{{\rm v_A}}{r}  = \frac{B_{\phi}}{\sqrt{4 \pi \rho} r}.
\end{equation}
The terms $N_T$ and $N_{\mu}$ are the thermal and composition components of the 
Brunt-V\"{a}is\"{a}l\"{a} frequency,
\begin{equation}
\label{NT}
N_T^2 = \frac{g\delta}{H_p}\left(\nabla_{ \rm ad} - \nabla_{ \rm rad}\right),
\end{equation}
and 
\begin{equation}
\label{Nmu}
N_{\mu}^2 = g\phi\left|\frac{\partial \ln\mu}{\partial r}\right|,
\end{equation} 
where $\nabla_{ \rm ad}$ and $\nabla_{ \rm rad}$ are the adiabatic and radiative gradients, 
$H_P$ is the pressure scale height, $g$ is the local gravity, $\mu$ is the mean 
molecular weight, $\delta = -\left(\partial \ln\rho/\partial \ln T\right)_{P,\mu}$, 
$\phi = \left(\partial \ln\rho/\partial \ln\mu\right)_{P,T}$, where $\rho$, T, and 
P are the local density, temperature and pressure, respectively. The thermal 
diffusivity is dominated by radiative transport, and is given by 
\begin{equation}
\kappa=\frac{16 \sigma T^3}{3\kappa_R \rho^2 c_P },
\end{equation}
 where $\gamma$ is the ratio of specific heats and $\kappa_R$ is the radiative opacity. 
The magnetic resistivity, $\eta$, is given by 
\begin{equation}
\label{eta}
\eta\approx 5.2 \times 10^{11}\frac{\ln \Lambda}{T^{3/2}} \, {\rm cm^2}\, {\rm s^{-1}},
\end{equation} 
\citep{spitzer06} where $\ln \Lambda$ is the Coulomb logarithm
\begin{equation}
\ln \Lambda \approx \begin{cases} -17.4 +1.5\ln T -0.5 \ln \rho & \,   T < 1.1\times10^5 \, \rm{K},\\
        -12.7+ \ln T -0.5 \ln \rho &\,   T > 1.1\times10^5 \, \rm{K}.
\end{cases}
\end{equation}
after translating into cgs units. In the current work, we assume the thermal viscosity is 
negligible compared to $\kappa$ and $\eta$ \citep{MBS04}.

\subsection{Instability and Growth Rate}
\label{instability}

The appropriate expressions for the instability criteria have terms
that depend on radial and on lateral gradients. The latter cannot be captured
in a one--dimensional code like MESA, so we address only the spherical 
radial components of the instability criteria.

\subsubsection{ST Instability}
\label{STinstab}

For the ST instability, a minimum initial magnetic field is required for growth. In our calculations 
of the ST dynamo process, we assume that this minimum field is present. In order for the overall 
ST dynamo process to work, however, a significant shear is required to overcome the effects of 
both thermal and compositional buoyancy. Following \citet{spruit02}, the shear condition for the 
ST process to operate may be expressed in our notation as
\begin{equation}
|q|> q_{\rm min}\equiv\left(\frac{N_{\rm lim}}{\Omega}\right)^{3/2}\left(\frac{\eta}{r^2 \Omega}\right)^{1/4},
\label{shearcond}
\end{equation}
where 
\begin{equation}
\label{ntotdef}
N_{\rm lim}^2=\left(\frac{\eta}{\kappa}\right)\max(N_T^2,0) +\max( N_{\mu}^2,0).
\end{equation}
Note that the sign of the shear is not important for the ST dynamo. This is because the only effect 
of the shear in the ST dynamo process is in winding up the poloidal field produced by the ST 
instability, and the field winding process depends only on the magnitude of the shear, not its sign.

\subsubsection{MRI Instability}
\label{MRIinstab}

Following \citet{BH91,BH98}, \citet{aki03}, \citet{MBS04} and \citep{KW14}, we can write the local instability 
criterion for the MRI in typical conditions in stars where the magnetic diffusivity $\eta$ is 
significantly smaller than the thermal diffusivity $\kappa$ as
\begin{equation}
\label{MRIcrit}
\left(\frac{\eta}{\kappa}\right)N_T^2 + N_{\mu}^2 + 2 q \Omega^2 < 0.
\end{equation}

In the absence of the Brunt--V\"{a}is\"{a}l\"{a} terms in Equation (\ref{MRIcrit}), the 
instability criterion for the MRI is simply $2 q \Omega^2 < 0$; that is, the system is 
unstable when $q < 0$, {\sl i.e.}, the angular velocity decreases outward. The component of the 
Brunt--V\"{a}is\"{a}l\"{a} frequency associated with composition gradients, $N_{\mu}^2$, 
is nearly always a stabilizing term in stars since the molecular weight almost always 
decreases monotonically outward. An exception arises in \S \ref{11} where we find 
a composition inversion with silicon overlying oxygen in the model with ZAMS mass 
of 11 \ms. The thermal component of the Brunt--V\"{a}is\"{a}l\"{a} frequency varies 
with the stellar structure. In convective regions, $N_T^2$ is negative and the
convective overturn promotes the MRI. In radiative regions, $N_T^2$ is positive and
this term will then tend to oppose the MRI instability. The thermal buoyancy term is,
however, diminished by diffusive effects. For small--scale perturbations, perturbed
fluid elements reach thermal equilibrium with the surroundings more quickly, thus
reducing thermal buoyancy and the associated stabilizing influence. The magnetic 
diffusivity will tend to promote stability because the tendency to amplify the field will 
diminish. Note that the latter is a very subtle effect, since that is the only indirect effect 
of the magnetic field. As long as the magnetic field is weak compared to the effects of 
rotation, $\omega_A << \Omega$, the instability criterion of Equation (\ref{MRIcrit}) 
does not depend on the strength of the magnetic field, one of the special properties 
of the MRI \citep{BH98}.

Note that that the precise definition of the ``reduced" $N_T$ depends on the particular instability. 
In the derivation of the MRI presented in \citet{KW14}, the term that we adopt in Equation (\ref{MRIcrit}), 
$N_{  T,\rm reduced}^2 = (\eta/\kappa) N_T^2$, corresponds to the instability criterion for the diffusive 
small--scale MRI. \citet{MBS04} present other instability criteria for which the appropriate reduced 
value is different. Although it is reasonable in the ambiance we explore in which $\eta << \kappa$, 
our expression would give unrealistically high estimates for the effects of buoyancy if 
$\eta>> \kappa$, giving $N_{T, \rm reduced} >> N_T$. 

While the growth rate in the ST mechanism depends on the field strength in a manner that
leads to predictions of the ratio of the resulting radial and toroidal field (\S \ref{Bfields}), 
the MRI is different in a fundamental way. If the field strength is below saturation, the growth 
rate of the MRI depends only on the shear, not on the strength of the magnetic field. In regions 
unstable to the MRI, the field should grow exponentially rapidly at the rate $q\Omega$. The 
growth rate for the MRI is likely to be much more rapid than that for the ST instability if 
the initial conditions correspond to a weak magnetic field, $\omega_A << \Omega$.

\subsection{Viscosity and Diffusion Coefficients}
\label{visc}

In MESA, all instabilities (including ST, Eddington--Sweet (meridional circulation), Goldrich--Schubert--Fricke)
and the MRI are treated as diffusive processes that diffuse angular momentum or species (mixing).
The net viscosity, $\nu$, is assumed to be the linear sum of the viscous diffusion coefficients 
$\nu_i$, corresponding to estimates of the diffusion coefficient for each individual process, $i$. 
Whether or not the diffusive effects associated with these various instabilities can truly be added 
in this simple linear way deserves deeper consideration, but that is beyond the scope of this work.  
Following \citet{spruit02}, the azimuthal stress, $S$, generated by the field produced by either 
ST or MRI can be related to an effective magnetic viscosity, ${\nu_{\rm mag}}$, by
\begin{equation}
\label{mag_stress}
S = \frac{B_r B_{\phi}}{4 \pi} 
= \rho q\Omega \nu_{\rm mag}. 
\end{equation}
This viscosity is explicitly an ``effective magnetic viscosity" that is determined by the global magnetic 
structure of the star and very specifically is not in any way related to the microphysics of ``molecular 
viscosity" in the star.

\subsubsection{ST Viscosity}
\label{STvisc}

For the ST process, the magnetic field components are first constrained by various physical arguments. The 
resulting prescriptions for $B_r$ and $B_{\phi}$ are then incorporated in Equation (\ref{mag_stress}) 
to evaluate the effective viscosity \citep{spruit99,spruit02}. The strength of the magnetic field 
components can be cast in a form in which the ST viscosity, $\nu_{\rm mag,ST}$, is treated as a variable (\S \ref{STfields}).

Our calculations for the magnetic viscosity, $\nu_{\rm mag,ST}$, corresponding to the ST mechanism 
are identical to those in \citet{HWS05} that are incorporated in MESA. The form of the equation 
for the effective ST viscosity depends on the signs 
of $N_T^2$ and $N_{\mu}^2$ and the strength of thermal diffusion, $\kappa$. In radiative regions, where 
both $N_T^2$ and $N_{\mu}^2$ are positive, we apply the effective viscosity calculated in Equations 
(34)-(37) of \citet{spruit02}. In semiconvective regions where $N_T^2<0$ and $N_{\mu}^2>0$, we 
apply Equations (6)-(9) of \citet{HWS05}. In thermohaline regions where $N_T^2>0$ and $N_{\mu}^2<0$, 
we apply Equation (36) of \citet{spruit02}, which corresponds to his ``Case 1." 

\subsubsection{MRI Viscosity}
\label{MRIvisc}

Dimensionally, $B \sim q\Omega r \sqrt{4 \pi \rho}$ (\S \ref{MRIfields}). Assuming the toroidal field 
to dominate, we can obtain a formal expression for the magnetic viscosity corresponding to the MRI 
by substituting this expression into Equation (\ref{mag_stress})
\begin{equation}
\label{viscosity1}
\nu_{ \rm mag, MRI} = \frac{B_r B_{\phi}}{4 \pi \rho |q| \Omega} = \left(\frac{B_r}{B_{\phi}}\right) |q| \Omega r^2,
\end{equation}
where the absolute value sign is used to ensure that $\nu$ is positive. We have 
not defined precisely what we mean by $B_r$ and $B_{\phi}$ in this context. We return to
this expression in \S \ref{MRIfields}.

To estimate the viscosity corresponding to the MRI, we have recourse to shearing--box
simulations. In accretion disks, the rotation is supersonic and the field produced by the MRI 
is limited to be less than the  value corresponding to equipartition with the local gas pressure,  
$\rho (\omega_{\rm A}r)^2\sim P_{\rm gas}$. Because the components of the magnetic 
field are turbulent, temporal and spatial averaging of simulation data is needed to obtain 
an accurate estimate of field components and their products. The normalized total magnetic 
pressure in a simulation can be expressed as  $<B^2>/(8 \pi P_0)$, where $P_0$ is the maximum 
magnetic pressure that can be produced by the MRI at saturation.  We adopt
$P_0=P_{\rm gas}$. If the radiation pressure, $P_{\rm rad}$, is 
significant, this expression should be replaced with $P_0=P_{\rm gas}+P_{\rm rad}$ 
to produce the correct normalization \citep{shi10}. In the calculations here,
the gas pressure and degeneracy pressure typically exceed the radiation pressure
by factors of at least several in the inner core. The addition of rotation and mixing
tends to increase $P_{\rm gas}/P_{\rm rad}$. We have neglected $P_{\rm rad}$
in our estimate of $\alpha$ in the current work. We argue that the corresponding 
normalization in the subsonic shearing conditions relevant to stars is 
$P_0=\rho (q \Omega r)^2$. We then assume that the appropriately normalized 
magnetic field components are the same in both accretion disks and in stars.

A stress efficiency parameter, $\alpha$, can then be defined as 
 \begin{equation}
\label{alphadef}
\alpha \equiv \frac{S}{P_0}= \frac{<B_r B_{\phi}>}{4 \pi P_0},
\end{equation}
where $<B_r B_{\phi}>$ is a suitable spatial and temporal average of the product of the
field components. As just argued, the normalization is $P_0 = P_{\rm gas}$ for an accretion 
disk and $P_0 = \rho (q \Omega r)^2$ for stars. We assume that the normalized parameter
$\alpha$ is the same in both disks and stars.

In local shearing--box simulations, the typical value of $\alpha$ is in the range 0.01 to 0.05 
\citep[][and references therein]{hawley11}. Global simulations may produce slightly larger 
values of $\alpha$, perhaps as large as 0.1 \citep[]{hawley11}. We adopt $\alpha = 0.02$ as
representative.
Using the right hand side of Equation (\ref{mag_stress}) and the definition of $\alpha$ in
the left hand side of Equation (\ref{alphadef}) yields an effective viscosity for the MRI of 
\begin{equation}
\label{viscosity3}
\nu_{ \rm MRI} =0.02 |q| \Omega r^2.
\end{equation}

We apply Equation (\ref{viscosity3}) for all values of $N_T^2$ and $N_\mu^2$ without 
modification as long as the instability criterion (\ref{MRIcrit}) is satisfied. The issue of how
the MRI works in semiconvective or thermohaline regions requires further work that is
beyond the scope of this paper. The prescription for viscosity is not modified
in semiconvective or thermohaline regions in the current work.

We have considered other physical conditions and associated prescriptions for the
effective viscosity of the MRI. \citet{spruit99} gives a prescription for the viscosity 
associated with the MRI (his Equation 31)
\begin{equation}
\label{viscosityS99}
\nu_{ \rm MRI} \sim 0.2 |q| \kappa \left(\frac{\Omega}{N_T}\right)^2.
\end{equation}
Spruit notes that this viscosity may be relatively small, but that his conclusion is preliminary 
pending numerical simulations of the non--linear development. This prescription was based on 
the assumption that conditions are held very near those corresponding to the onset of 
the linear instability. It is not clear to us that the system under consideration will maintain 
this marginal condition. We have, rather, invoked estimates corresponding to something 
like saturation as revealed by simulations. We have, however, run one 15\ms\ model 
(\S \ref{15}) with the prescription of Equation (\ref{viscosityS99}) and find that it can be 
comparable to, or even exceed the prescription we adopt in Equation (\ref{viscosity3}). 

Another concern is that the instability criterion, Equation (\ref{MRIcrit}), specifically
invokes the destabilizing effect of thermal diffusion. The question arises as to 
whether or not the effectiveness of the MRI in providing a viscosity is also limited
by the constraint of significant thermal diffusion. Since the growth time of the magnetic
field is given by the shear, the Maxwell stress is of order
the Reynolds stress, $S \sim \rho \ell_r \ell_{\phi} \sigma^2$, where $\ell_r$ and
$\ell_{\phi}$ are characteristic length scales in the radial and azimuthal directions
and $\sigma \sim q \Omega$. If to maintain the growth of the MRI,
the length scales are restricted to be sufficiently small that thermal diffusion is
active, then the effectve stress and associated viscosity might be also limited. 
Suppose, for example, that $\ell_{\phi} \sim r$ ( or a pressure scale height), but 
that $\ell_r$ is restricted by the condition of effective thermal diffusivity. The
latter could be expressed by writing $k^2 \kappa \sim N$ where $k$ is the wavenumber
of the maximally destabilized mode and $N$ is the Brunt-V\"{a}is\"{a}l\"{a} frequency. 
The constraint on the length scale can thus be expressed as $\ell_r \sim \sqrt{\kappa/N}$
and the stress as $S \sim \rho q^2 \Omega^2 r \sqrt{\kappa/N}$. The associated viscosity
would then be 
\begin{equation}
\label{viscosityS14}
\nu \sim |q| \Omega r \sqrt{\kappa/N}, 
\end{equation}
smaller than we adopted in
Equation (\ref{viscosity3}) by a factor of roughly $\sqrt{\kappa/N}/r$. We have been somewhat
loose in this discussion with the exact nature of the Brunt-V\"{a}is\"{a}l\"{a} frequency, N.
The relevant choice would seem to be the thermal component, $N_T$, since this
sets the relevant buoyancy timescale and, in the absence of thermal destablizing 
effects, dominates the composition term. Conditions for which $N_T^2  <  0$ will
be convective and the effective convective dynamic viscosity would then dominate
other effects. We have adopted the prescription of Equation (\ref{viscosityS14}) in a 
model of a 15 \ms star (\S \ref{15}) in regions that are unstable to the MRI
and for which $N_T^2  >  0$ with no other magnetic effects. We find, as expected, 
that the viscous mixing and transport effects of the MRI are rather small. 

Given instability according to Equation (\ref{MRIcrit}), the question becomes whether
the prescription of Equation (\ref{viscosity3}) or Equation (\ref{viscosityS14}) best
describes the effective viscosity as the field grows toward saturation. In unstable conditions, 
there will be a most--rapidly growing mode of wave number, $k_{mrg} {\rm v_A} \sim \Omega$ 
with a corresponding length scale 
\begin{equation}
\label{mrg}
\ell_{mrg} \sim \frac{{\rm v_A}}{\Omega} \sim \rho^{-1/2} \frac{B}{\Omega},
\end{equation}
neglecting factors of order unity. Suppose the MRI sets in with a small ambient
field such that $\ell_{mrg} < \ell_r \sim \sqrt{\kappa/N}$. The field will then grow until
these two length scales are comparable, corresponding to a field strength of order
\begin{equation}
\label{Bthermal}
B \sim \left(\frac{\kappa \rho}{N}\right)^{1/2} \Omega. 
\end{equation}
It is not clear that this condition will suppress further field growth with this characteristic 
wave number, and even if it does, there will be perturbation due to MRI turbulence that 
will be of larger wave number and smaller length scale that can continue to grow in an 
unstable environment, albeit at a slower rate. Similar perspectives pertain even if the 
most rapidly-growing mode has a characterstic length larger than the thermal diffusive 
length scale even at the onset of instability. As long as some mode grows on a time 
scale that is short compared to the evolutionary times in the star, it seems that the field 
should continue to grow in strength, and that the only natural limit is that of saturation 
with ${\rm v_A} \sim q \Omega$.  This is basically the condition that underlies 
Equation (\ref{viscosity3}). 

In possibly analogous situations, double--diffusive instabilities that might yield 
sufficient perturbations to provide a torque are rendered ineffective because of
associated small scale turbulence that prevents effective radial coupling in the
shear flow \citep{deniss10}. Even at saturation, this might affect the effective viscosity
of the MRI. For the reasons described here, we have presented results using
Equation (\ref{viscosity3}) for the MRI viscosity based on extant numerical MRI
simulations but recognize that there are issues of physics here that require greater study.

\subsubsection{ST and MRI Diffusion Coefficients}
\label{diff}

We now discuss the species mixing produced by each instability.  The net diffusion coefficient 
for mixing, $D$, is determined  by linearly adding the diffusion coefficients $D_i$ corresponding 
to each process weighted by an efficiency factor $f_{c,i}$ which we discuss later in this section. 
For all of the hydrodynamic instabilities, $\nu_i=D_i$. For the ST mechanism, the mixing 
is produced by the effective magnetic resistivity rather than the effective magnetic viscosity and 
again depends on the signs of  $N_T^2$ and $N_{\mu}^2$ and the strength of thermal diffusion, 
$\kappa$. Our prescriptions are identical to those in \citet{HWS05} that have been incorporated 
in MESA. In radiative regions, we apply the effective resistivity given by Equations (41)-(43) of 
\citet{spruit02}. In semiconvective regions, we apply Equations (6)-(9) of \citet{HWS05}.  
In thermohaline regions, we apply Equation (43) of \citet{spruit02}, which corresponds to his ``Case 1".

In the rubric of the MRI, the quantities $D_{ \rm MRI}$ and $\nu_{ \rm MRI}$, are expected to be 
about the same amplitude, at least under conditions of marginal stability \citep{M09}. Models of 
the turbulent mixing associated with the MRI give a range of values of the ratio 
$D_{ \rm MRI}/\nu_{ \rm MRI}$. The presence of an initial vertical magnetic field may decrease 
$D_{ \rm MRI}$ relative to $\nu_{ \rm MRI}$ \citep{JKM06}, but the radial diffusion coefficient 
remains within a factor of three of $\nu$ \citep{Arm11}. In the present work we thus take 
$D_{ \rm MRI}=\nu_{ \rm MRI}$ from Equation \ref{viscosity3} in radiative, semiconvective,
and thermohaline regions. We note that the model with ZAMS mass of 11 \ms\ forms composition
inversions that might trigger thermohaline instability, but we do not consider that in this paper.

There are various efficiency factors related to the mixing process. Following \citet{HLW00}, 
in MESA $f_c$ is taken to be unity for convection and semi--convection and is taken to be a 
constant, $f_c = 1/30$, for the other processes. A second parameter, $f_{\mu} = 0.05$,
is used to weight the $\mu$--gradients in the individual terms. \citet{HLW00} calibrated $f_c$
and $f_{\mu}$ by comparing observed surface abundances of nitrogen in lower mass, solar--type 
stars with model results based only on hydrodynamic instabilities. It is not completely clear that 
this calibration also applies to higher mass stars and when invoking magnetic instabilities, but 
this value has also been used in other studies invoking the ST instability and comparison with 
surface nitrogen abundances in more massive stars \citep{Brott11,Ekstrom12,YDL12}. We adopt 
it for the ST mechanism on the grounds of consistency with other, similar work.  

In contrast, there is direct evidence from MRI simulations, as described above, that the mixing 
and diffusion coefficients are nearly equal. Due to the small scale length of the most rapidly--growing 
MRI modes, adding radial stratification that would be present in stars but is not present in 
those simulations might then make little difference to the results. Given these considerations, 
the large range in $D_{ \rm MRI}$, the large value of $D_{ \rm MRI}$ when the MRI is active, and 
the intrinsic uncertainties in the formulation and implementation of the mixing of species, 
we set the condition $f_{ \rm c,MRI} = 1.0$, but adopt $f_{\mu, MRI} = 0.05$. We ran one of our
fiducial models of ZAMS mass of 15 \ms\ with $f_{ \rm c,MRI} = 1/30$, the coefficient adopted for
the ST mechanism. There were small quantitative but no qualitative differences compared to the 
model with $f_{ \rm c,MRI} = 1.0$. The most distinct difference was that the model with
$f_{ \rm c,MRI} = 1/30$ showed a more ragged composition distribution compared to the
smoother distributions found with $f_{ \rm c,MRI} = 1.0$.

The mixing and diffusion associated with the MRI proceed on similar timescales of order
\begin{equation}
\label{mixing1}
\tau_{ \rm mri} \sim \left( \frac{\ell^2}{D_{ \rm mri}} \right) \sim \left( \frac{D_{ \rm mri}}{{\rm v}^2} \right)
\end{equation}
where $\ell$ and ${\rm v}$ are characteristic length and velocity scales for the mixing. For the MRI, 
the length scale of the mixing is plausibly less than the pressure scale height, $H_p$ and, because 
both ST and the MRI are magnetic effects, the characteristic velocity associated with either of them 
is likely to be restricted to ${\rm v} < {\rm v_A}$. With these limits and with Equation \ref{viscosity3}, 
Equation \ref{mixing1} can be recast in the form
\begin{equation}
\label{mixing2}
\tau_{ \rm mri}~  \ltsim \frac{H_p^2}{0.02 |q| \Omega r^2} = 1000~{\rm s} \frac{H_{p,9}^2}{D_{15}},
\end{equation}
where $H_{p,9}$ is the pressure scale height in units of $10^9$ cm and $D_{15}$ is the diffusion
coefficient in units of $10^{15}$ cm$^2$ s$^{-1}$, a characteristic value when the MRI is active.
The mixing is thus potentially very rapid. The MESA time steps are of order 100 years at the end 
of core helium burning in the 15\ms\ model, so they are long compared to the diffusion timescale 
given in Equation (\ref{mixing2}). Our treatment of MRI mixing thus considers it to be ``instantaneous" 
in that phase, analogous to assuming ``instantaneous" mixing by convection in fully--efficient convective 
regions in a more traditional context. By the onset of core collapse in that model, the MESA time steps 
decline to be of order 1 second. The MRI mixing might thus be resolved in that limit. The mixing may 
change the structure in a way that mutes the mixing by altering the gradient in angular velocity. This 
is a complex problem. In this work we have not attempted to specifically resolve 
the variations of the angular momentum per unit mass, the angular velocity and the composition 
profile on the short timescales indicated by Equation (\ref{mixing2}) nor to  determine how these 
functions vary as parameters are altered. Rather we have chosen to show discrete intermediate 
stages and the final core mass and composition structure as integral measures of all these 
complex effects. We leave more detailed studies for future work.

\subsection{Magnetic Fields}
\label{Bfields}

\subsubsection{ST Saturation Fields}
\label{STfields}

In the range of length scales bounded below by magnetic diffusion and above by stratification, 
\citet{spruit02} argued that the growth rate for the ST mechanism is $\omega_A^2/\Omega$.
The condition for field saturation for the ST mechanism is that the growth rate is balanced by 
magnetic diffusion. Because the growth rate depends on the field strength, prescriptions can
be written for the toroidal and radial field strengths that depend on the rotation, the shear, the 
buoyancy terms, and the thermal diffusivity. \citet{spruit02} gives prescriptions for the radial 
and toroidal fields corresponding to the ST mechanism in the limits where the thermal diffusion 
can be ignored and where it dominates. In the former case, the appropriate expressions are
\begin{equation}
B_{\phi} \approx \frac{\left( 4 \pi \rho \right)^{1/2} q r \Omega^2}{N_T}
\end{equation}
and
\begin{equation}
B_r \approx B_{\phi}  q \left(\frac{\Omega}{N_T}\right)^2.
\end{equation}

An effective magnetic ST viscosity, $\nu_{\rm mag,ST}$ can then be computed from the field
components (\S \ref{STvisc}). As a computational convenience, \citet{HWS05} give a general 
prescription for the ST magnetic field components in terms of $\nu_{\rm mag,ST}$ in their 
Equations (11) and (12) as
\begin{equation}
\label{BphiST}
B_\phi^4=16\pi^2\rho^2 \nu_{\rm mag,ST}q^2\Omega^3 r^{2},
\end{equation}
\begin{equation}
\label{BrST}
B_r^4=16\pi^2\rho^2 \nu_{\rm mag,ST}^3q^2\Omega r^{-2}.
\end{equation}
We use this prescription to calculate the magnetic field from the effective viscosity. The ratio 
of the squares of the magnetic field componenets is then given by
\begin{equation}
\label{STratio}
\frac{B_r^2}{B_{\phi}^2}=\frac{\nu_{\rm ST}}{r^2 \Omega} .
\end{equation}
This ratio is always much smaller than unity, so the toroidal field produced by the ST 
mechanism is dominant and the resulting magnetic viscosity relatively modest.

\subsubsection{MRI Saturation Fields}
\label{MRIfields}

For the subsonic flow conditions present within a star, the saturation field for the MRI can be 
estimated to order of magnitude by using the saturation condition $\omega_A \sim q \Omega$ 
\citep{BH98,vishniac09}. Assuming that the toroidal field dominates, we therefore have
 \begin{equation}
\label{alfvenmri}
B_{\phi} \sim q\Omega r \sqrt{4 \pi \rho}.
\end{equation} 

A somewhat more precise estimate of the toroidal field strength and an estimate of
the radial field for the MRI can be based on numerical simulations of accretion disks. 
To avoid cancellations during the averaging, we make the identifications 
$\overline{B_r} = \sqrt{<B_r^2>}$ and $\overline{B_\phi} = \sqrt{<B_\phi^2>}$, 
where the brackets represent temporal and spatial averaging (\S \ref{MRIvisc}). 
In shearing box accretion disk simulations, $\overline{B_\phi}^2/(8 \pi P_0) \approx 0.08$ 
\citep[][and references therein]{hawley11}, where the normalization is $P_0 = P_{\rm gas}$ 
for an accretion disk and $P_0 = \rho (q \Omega r)^2$ for stars (\S \ref{MRIvisc}). The 
ratio of the squares of the two components from simulations is $\overline{B_r}^2/\overline{B_\phi}^2 
\approx 0.1$. Global simulations indicate a somewhat larger value, $\overline{B_r}^2/\overline{B_\phi}^2 
\approx 0.2$ \citep{hawley13}, but in the absence of a converged estimate from such simulations 
we use shearing--box estimates for the magnetic field for consistency. A possible concern is 
that the saturation field of the MRI and the resulting shear might have a strong dependence on 
the wavenumbers of the unstable modes \citep{DSP10}, but see \citet{vishniac09}.

Using the above values, we adopt an estimate for the toroidal field of 
\begin{equation}
\label{bphi}
\overline{B_\phi} \approx 0.40 q\Omega r \sqrt{4\pi \rho},
\end{equation}
and an estimate of the radial field of
\begin{equation}
\label{br}
\overline{B_r} \approx  0.32 B_{\phi}\approx 0.13 q\Omega r \sqrt{4\pi \rho}.
\end{equation}
Note that in this prescription for the MRI, the ratio of the radial to toroidal field, $\approx 0.3$,
is constant and much larger than the corresponding ratio for the ST mechanism. 
This has implications for the corresponding magnetic viscosity (\S \ref{MRIvisc}).

Unlike the prescription for the magnetic viscosity of the ST mechanism, our assignment of the 
magnetic viscosity of the MRI based on simulations does not require a prescription for the
ratio of $B_r$ to $B_{\phi}$, nor vice versa; nevertheless, these factors are generically related. 
Using Equation (\ref{bphi}) for $B_{\phi}$ and Equation (\ref{br}) for $B_r$ in Equation 
(\ref{mag_stress}) yields an effective value of $\alpha = 0.05$, a formal discrepancy of a factor 
of 2.5 with respect to the value we adopt for $\alpha$ in Equation (\ref{viscosity3}). We ascribe 
this discrepancy to differences in the averaging procedure in the numerical simulations, such that 
$<B_r B_\phi> \ne \overline{B_r} \overline{B_\phi}$. There is probably some cancellation of 
opposite signs of $B_r$ and $B_{\phi}$ at different locations in the calculation of the stress that 
are not reflected in calculating the mean squared components. The value of $\nu_{ \rm MRI}$ 
that we adopt in Equation (\ref{viscosity3}) roughly corresponds to $(\overline{B_r}/\overline{B_\phi})^2 
= 0.1$. Comparison to the corresponding ratio for the ST mechanism from Equation (\ref{STratio}),
shows that where it is active, the MRI has a much larger effective magnetic viscosity than the ST mechanism.

\section{Results}
\label{results}

In the present work, we have used the stellar evolution code MESA, version 5456 (Paxton et al. 2011;
2013) to evolve models of massive stars. Standard mass--loss rate prescriptions appropriate for 
massive stars were employed \citep{deJager88, vink01}. The effects of rotation on mass loss are 
treated using the approximation presented in \citet{HLW00}. For cases approaching the 
critical angular frequency, the mass loss rate is limited by the thermal timescale following the 
prescription of \citet{YWL10}. We used the Helmholtz equation of state \citep{TS00} that includes 
the contributions from e$^{+}$+ e$^{-}$ pairs and the ``approx21" nuclear reaction network \citep{T99}.

Rotation in MESA is treated using the prescriptions of \citet{HLW00} and \citet{HWS05} that include 
many relevant hydrodynamical instabilities that affect the mixing of chemical species and angular 
momentum transport (Eddington--Sweet (ES) meridional circulation, the dynamical and secular 
shear instabilities and the Solberg--Hoiland and Goldreich--Schubert--Fricke (GSF) instabilities). 
MESA has also the capability of including the effects of magnetic fields on angular momentum 
transport and mixing of species based on the ST prescriptions from \citet{spruit99} and \citet{spruit02}. 
MESA calculates the ratio $\eta/\kappa$. This calculation is done taking account of appropriate 
prescriptions for degenerate matter. Typical values in the central regions of our massive star models 
are $\eta/\kappa \approx 10^{-12}$, so the muting of buoyancy stability is appreciable. 

We explored the effects of the MRI for a range of ZAMS masses, 7, 11, 15, and 20 \ms,
all at solar metallicity. In the cases discussed below, ``depletion" is defined as a central mass 
fraction of the relevant element becoming less than $10^{-4}$. All the models were run incorporating 
the default treatment in MESA for convection, semi-convection, dynamical and secular shear 
instabilities, ES circulation, and the Solberg--Hoiland and GSF instabilities. We found that the ES 
circulation dominated the non--magnetic, rotationally--induced processes. In the plots given below, 
we present only the diffusive and mixing effects of the ES circulation. 

The model with 7 \ms\ was run with only the MRI magnetic physics. The other three ZAMS 
masses were run for the five cases, with no rotation (``no-rot" models), rotation with all the 
standard mixing and diffusive instablities but no magnetic mixing or diffusion effects 
(``rot-none" models), rotation with the standard effects plus the ST prescription for mixing of 
species and transport of angular momentum (``rot-st" models), rotation with the standard effects 
plus the MRI prescription for mixing of species and transport of angular momentum (``rot-MRI" models), 
and rotation with the standard effects plus both ST and MRI prescriptions activated (``rot-mrist" models). 
Note that while the two magnetic effects interact in the simulation, they are invoked with separate 
prescriptions, viscosities, and diffusion coefficients, rather than being treated as fundamentally related 
in terms of common linear instability and subsequent growth of the instability. 

For each ZAMS mass, we elect an initial surface equatorial velocity of 206 \kms\ 
\citep{HWS05}. For the model with ZAMS mass of 15 \ms, this represents one of 
the MESA test problems that has been verified and benchmarked against other codes. 
This initial value of the velocity represents a characteristic rotation velocity for massive stars
and has been used as a fiducial value by many authors \citep{HLW00, HWS05, Brott11}.
In the future, a more thorough study would involve a variation of this parameter, but 
for this preliminary study we adopt this single representative value.

In regions where the MRI is active, the value of log $D_{ \rm MRI}$ varies substantially. When 
log $D_{ \rm MRI}$ is low, the MRI is only marginally unstable at this specific place and time. Other 
processes, mainly meridional circulation and regular convection, dominate mixing in the regions 
where log $D_{ \rm MRI}$~ 9-12. Only at values close to log $D_{ \rm MRI}$ $\sim$15-20 is the 
MRI prominent thanks to the strong shear, especially at core boundaries. 

The model with 7 \ms\ was chosen to explore whether or not the MRI might change the 
boundary between degenerate CO and ONeMg core evolution. The model with 11 \ms\ 
falls in a range where the evolution is very sensitive to ZAMS mass and treatment of
physics, is associated with electron--capture core collapse in classic treatments
\citep{Miyaji80}, and may fall in the range for which searches have identified red--giant 
progenitors \citep{smartt09}. The models with 15 and 20 \ms\ are in the range
where iron--core collapse occurs and perhaps at the upper end of explosions for
which red--giant progenitors are clearly identified. We adopted the 15 \ms\ model
as our fiducial model and explore its nature in somewhat more depth in \S \ref{15}.

Examination of these models shows that while the MRI is suppressed in the earliest stages 
of the evolution, the instability criterion of Equation (\ref{MRIcrit}) is satisfied in portions 
of the structure at more advanced stages. In the absence of the effects of the ST instability, 
the MRI alone can result in some mixing and homogenization of the structure and some 
transport of angular momentum that is different from the standard treatment, given the 
prescriptions we have adopted here. One result is that the MRI, in the absence of ST effects, 
yields a somewhat smaller iron core than the basic non--rotating model. Without ST effects, 
the MRI in conjunction with standard processes can lead to rather small rotation rates 
of the iron core. 

\subsection{7 \ms\ Model}

\begin{figure}
\begin{center}
\includegraphics[angle=-90,width=16cm]{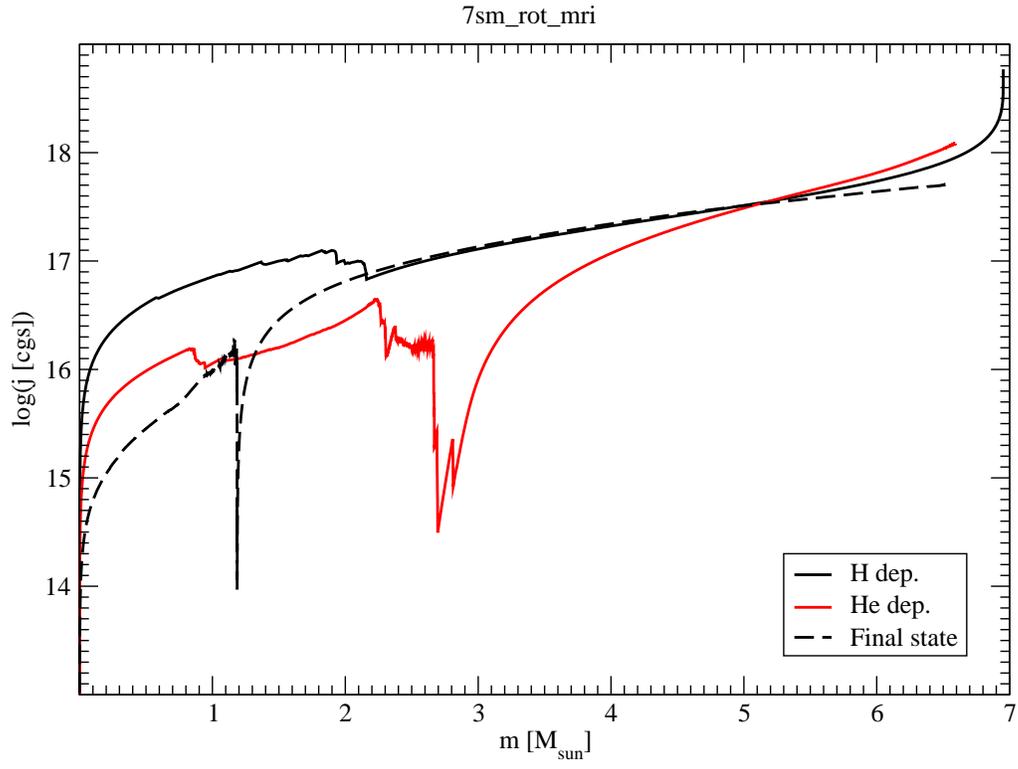}
\caption{Distribution with respect to mass of the specific angular momentum in the model with
ZAMS mass of 7 \ms\ at the end of hydrogen burning, at the end of helium burning, and in
the final model with a degenerate O/Ne core for the model with MRI, but not ST, active. See the online version for color.}
\label{7j} 
\end{center}
\end{figure}

We did not investigate the model with ZAMS mass of 7 \ms\ in the detail of the more 
massive models, but only investigated a model with the MRI magnetic physics. 
This model evolved to form a degenerate core of intermediate mass elements with a 
central density of $3\times10^7$ \gcm\ and a central temperature of $3\times10^8$ K, 
at which point the evolution was artificially halted. The final temperature profile 
showed a temperature inversion due to neutrino losses. Figure \ref{7j} gives the 
distribution of the angular momentum per unit mass, j, at the phase of hydrogen 
depletion, at the phase of helium depletion, and in the final model. The steep drop 
in j at about 2.7 \ms\ at helium depletion and at 1.2 \ms\ in the final model are 
due to the viscous action of the MRI. The inner core is not spun down drastically in
the final model and the angular momentum in the outer envelope is rather modest.

\begin{figure}
\begin{center}
\includegraphics[angle=-90,width=16cm]{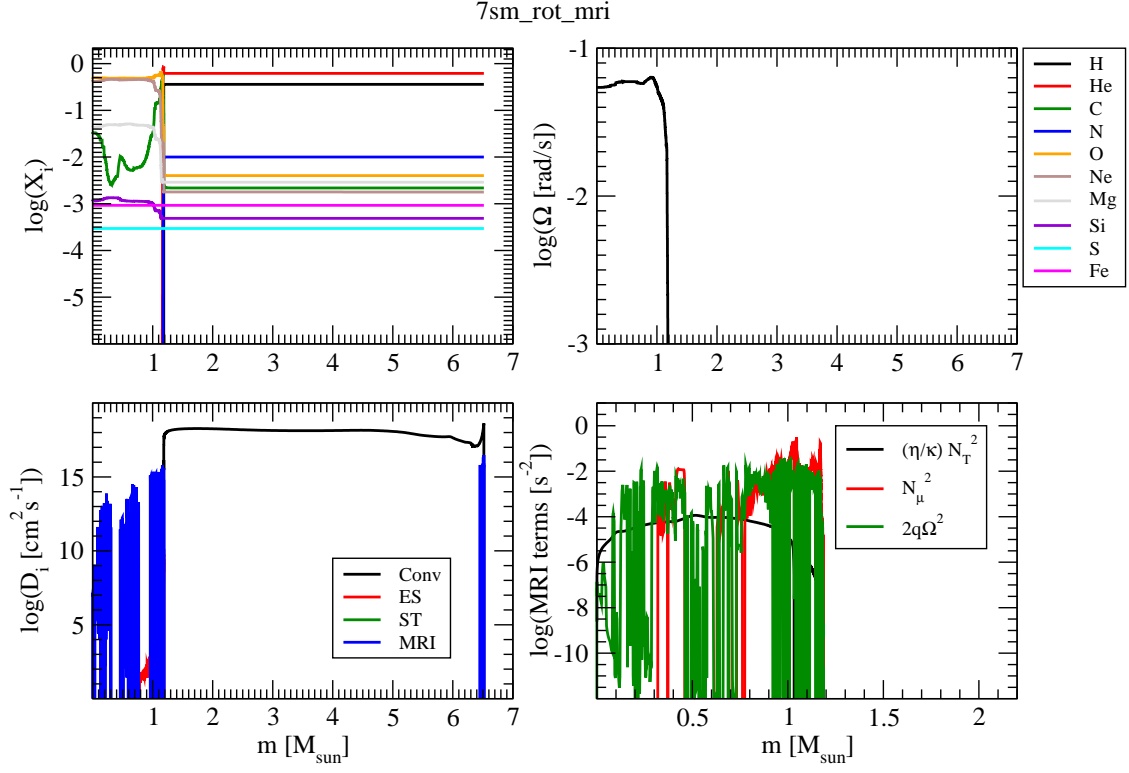}
\caption{Distribution with respect to mass  in the final model of
ZAMS mass of 7 \ms\ of the composition (upper left), the angular velocity, $\Omega$ (upper right),
the components of the diffusion coefficient (lower left), and the components of the MRI 
instability criterion (lower right) for the model with MRI, but not ST, active. See the online version for color.}
\label{7fs} 
\end{center}
\end{figure}

For the final model of 7 \ms, Figure \ref{7fs} gives the composition distribution (upper left), 
the distribution of the angular velocity, $\Omega$, (upper right), the diffusion coefficients 
corresponding to thermal convection (``conv"), ES circulation, and the MRI (lower left) and 
the components of the MRI instability criterion,$(\eta/\kappa) N_T^2$, $N_{\mu}^2$ and 
$2 |q| \Omega^2$ (lower right). Note that the suppressed thermal component 
of the Brunt--V\"{a}is\"{a}l\"{a} frequency is generally negligible throughout the inner core. 
This component is negative in the outer convective envelope and hence not plotted, but would 
slightly promote the MRI there in the presence of any shear. In the core, the destabilizing component, 
$2q\Omega^2$, frequently dominates over the stabilizing term, $N_{\mu}^2$. The activity of 
the MRI in terms of its dominance of the diffusion coefficients in the inner core is clear. That 
core is in nearly solid body rotation in the model (upper right panel of Figure 2). 

Of greatest interest is the final composition distribution. In this model, the core was composed 
essentially half each by mass of oxygen and neon. The carbon was nearly burned away, with a 
mass fraction of substantially less than 0.01 through most of the core. The magnesium mass 
fraction was about 0.05. In this model with an active MRI, the final core more closely resembles 
that expected to undergo electron--capture induced core collapse than degenerate carbon 
ignition with subsequent deflagration and detonation. In practice, such a star, if single, is 
likely to lose its hydrogen envelope to form a planetary nebula, but if such a star were in a 
binary system it might undergo a later evolution driven by mass accretion. The question of 
whether or not the small remaining carbon would affect the evolution is a very interesting 
one we postpone for later investigation.

\subsection{11 \ms\ Model}
\label{11}

The models with ZAMS mass of 11 \ms\ fall in a range that is notoriously sensitive to treatment 
of input physics. All these models ran very slowly toward the end, and none were run to a truly
final end point. The models were artifically halted when the evolution became unacceptably slow, 
an unfortunately subjective criterion. The result was that models with different parameters 
were run to somewhat different stages, making the intercomparison of models cumbersome.
The models were stopped at the following densities in units of $10^6$ \gcm\ and times in units 
of $10^7$ yr: non--rotating, 64, 1.936; rotating but no magnetic effects, 1,5, 1.965; ST only, 
1.6, 1.978, MRI only, 49, 2.309; ST and MRI, 203, 2.085. 

\begin{figure}
\begin{center}
\includegraphics[angle=-90,width=16cm]{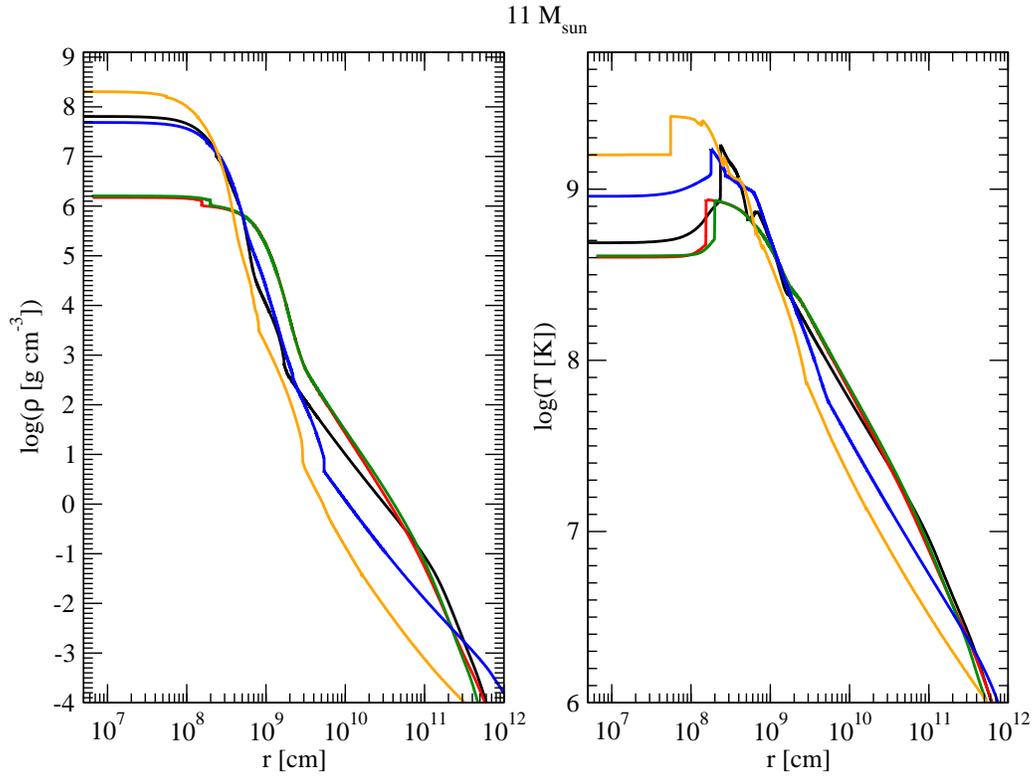}
\caption{Radial distributions of density and temperature for the ``final"
models corresponding to ZAMS mass of 11 \ms\ for the cases with no rotation (black), rotation but no magnetic 
effects (red), ST but not MRI (green), MRI but not ST (blue), and with both ST and MRI active (orange). 
These models were halted artificially, see text. See the online version for color.
}
\label{11structure} 
\end{center}
\end{figure}

Figure \ref{11structure} shows the  
density and temperature structures at these epochs. The decreased core temperatures reveal 
the effect of neutrino cooling. The models with no rotation and with MRI magnetic effects only 
give similar density profiles but somewhat different core temperatures. The models with rotation 
with no magnetic effects and those with ST only show very similar density and temperature profiles, 
perhaps because they were both halted at rather lower densities and earlier times. The model 
with both ST and MRI essentially finished core oxygen burning and gave the most extreme core 
densities and temperatures and the smallest inner, cooler core. It is not clear why this model
was able to proceed further in its evolution, but the extra mixing apparently allowed the
model to more smoothly converge for a longer time. 

\begin{figure}
\begin{center} 
\includegraphics[angle=-90,width=16cm]{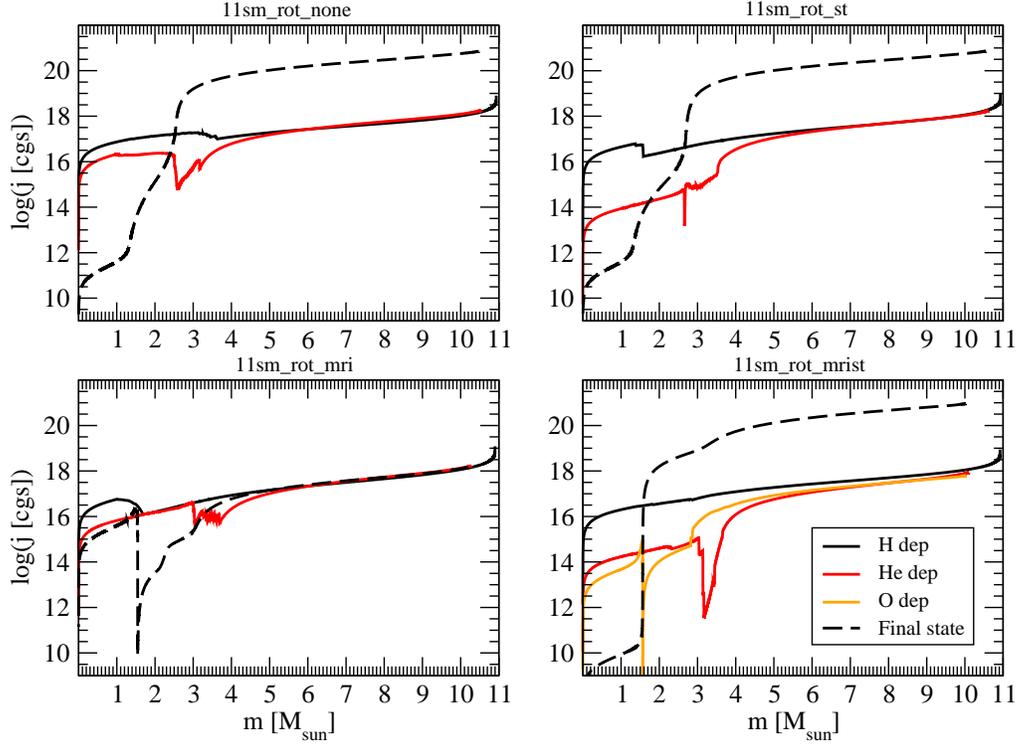}
\caption{Distribution with respect to mass in the  models with ZAMS mass of 11 \ms\ of the 
specific angular momentum at the end of hydrogen burning (black solid line), helium burning 
(red line), oxygen burning (orange line), and for the ``final" model (black dashed line) for the cases 
with rotation but no magnetic effects (upper left), ST but not MRI (upper right), MRI but not ST
(lower left), and with both ST and MRI active (lower left). These models were halted artificially, see text.
None reached core collapse and only the model with both ST and MRI active completed oxygen burning. 
See the online version for color.}
\label{11j}
\end{center}
\end{figure}

Figure \ref{11j}  shows the ``final" respective distributions of angular momentum per unit mass
for the 11 \ms\ models. The model MRI magnetic effects alone does not yield the strong spin--down 
of the core compared to other effects, but does show a spin--down of the matter just beyond the 
core (refer to Figure\ref{7j}). This model has an envelope with relatively small angular momentum, 
suggesting that the core has not transferred angular momentum outward as have the other models. 
It appears that the presence of the MRI, but not ST, is inhibiting the outward angular momentum 
transport that characterizes even the model with only the generic transport effects in the upper 
left panel. The model with both ST and MRI does yield a slowly--rotating core after oxygen 
burning.

\begin{figure}
\begin{center}
\includegraphics[angle=-90,width=16cm]{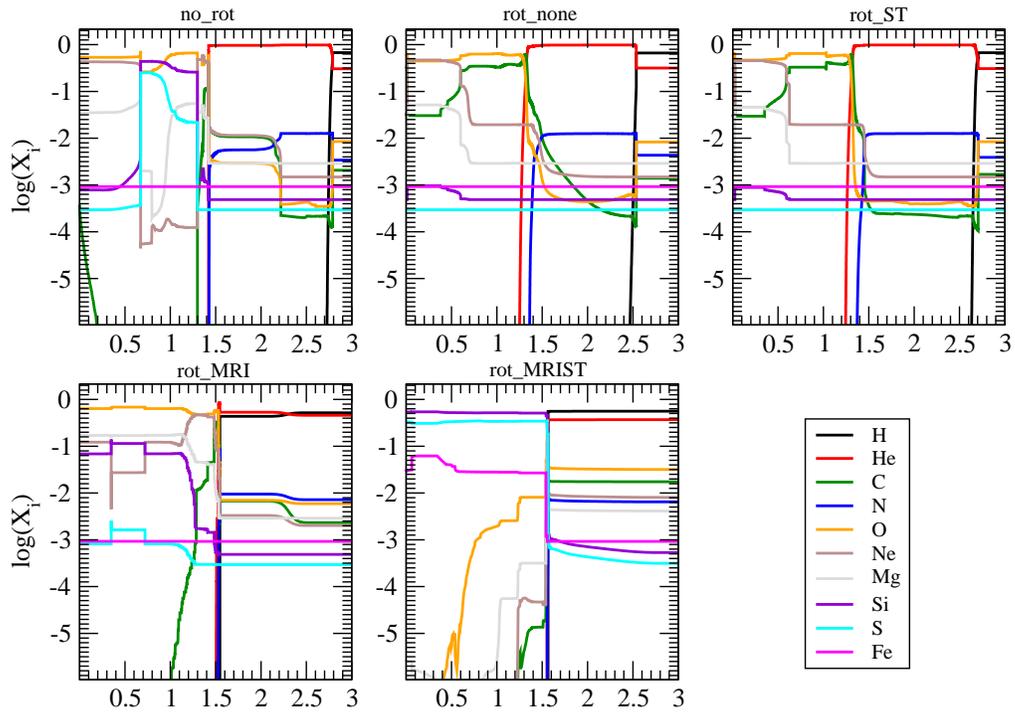}
\caption{Distribution with respect to mass in the model with ZAMS mass of 11 \ms\ of the ``final"
composition of the various models of Figure \ref{11structure}. See the online version for color. }
\label{11comp} 
\end{center}
\end{figure}

Figure \ref{11comp} shows the final composition profiles for the 11 \ms\ models with no rotation, 
rotation but with the magnetic effects suppressed, with ST only, with MRI only, and with both ST 
and MRI implemented. The non--rotating model developed a neutrino--cooled ONe core, 
but with an overlying silicon-rich layer in which oxygen and sulfur had equivalent abundances
after shell burning there. This structure may be unstable to thermohaline mixing \citep{mocak11}. 
The model with rotation but no magnetic effects resembled that with ST alone, both of which produced 
cores of oxygen and neon with an overlying layer of carbon and oxygen. The model with MRI 
alone produced a very oxygen--rich core with rather small traces of magnesium and other elements. 
The model with both ST and MRI enabled produced a nearly homogeneous Si/S core with oxygen 
nearly burned out and iron growing in abundance. For both the MRI model and the model with 
both ST and MRI, the core interior to the helium mantle has a mass of 1.5 \ms, significantly 
above the Chandrasekhar limit for a mean molecular weight per electron of 2. These models cannot 
support a degenerate core and seem destined to proceed to collapse of some sort, most 
likely to iron--core collapse.

This mass range merits much further detailed study, but the suggestion is that magnetic 
effects can promote the formation of an iron core in a mass range that would otherwise 
be predicted to lead to O/Ne/Mg cores and electron--capture induced collapse.

\subsection{15 \ms\ Model}
\label{15}

We adopted the model with ZAMS mass of 15 \ms\ as our fiducial model and present here
a more detailed exposition of its properties. All 15 \ms\ models proceeded through the
end of core Si burning, defined when $X_{ \rm center,Si} < 10^{-4}$ \citep{HWS05}. The 15 \ms\
models were halted by the flag in MESA indicating the onset of the phase of dynamical collapse 
of the iron core. In each model, the outer edge of the iron core is defined by the condition 
$X_{ \rm Fe} = 0.5$. Table \ref{models} gives for each of the four assumptions concerning rotating 
models the final mass of the model, the final mass of the iron core, the final equatorial 
velocity of the outer edge of the model, and the final equatorial velocity at the edge of the iron core.

Figure \ref{15structure} gives the density, temperature, pressure, and mean molecular weight
distributions as a function of radius in the 15 \ms\ models at the end of the calculation. 
The differences in the models with no rotation, rotation effects but no magnetic effects, 
ST only, MRI only, and with both ST and MRI invoked are rather small. The most noticeable 
differences are in the composition distribution that results from the different degrees of mixing. 

Figure \ref{15rotation} gives the distributions of angular velocity, $\Omega$, and the equatorial velocity
at the end of the simulation of the 15 \ms\ models for the cases with no magnetic effects, for ST 
only, for MRI only, and for both ST and MRI prescriptions invoked. For the model with only 
the magnetic effects of the ST mechanism, the iron core has been rendered nearly irrotational. There is still 
some remnant angular momentum, but it is very small, in agreement with the results of \citet{HWS05}.

Figure \ref{15B} gives the distributions of the estimated magnetic fields for the model of 15 \ms\ 
just prior to core collapse for the model where only the ST is active using Equations (\ref{BphiST})
and (\ref{BrST}) and for the model where only the MRI is active using Equations (\ref{bphi})
and (\ref{br}). The prescription for the ST fields yields modest toroidal field strength, $\sim
10^8$ to $10^9$ G and a radial component that is typically a factor $\sim 10^4$ times smaller
than the toroidal component. Both of these factors contribute to a rather modest magnetic
viscosity (Equation \ref{mag_stress}). Although it is sparsely distributed, the peak toroidal
field is much larger for the MRI, $\sim 10^{12}$ to $10^{13}$ G, and the radial field is
a significant fraction of the toroidal field (0.32 in this work). These factors contribute to
a larger magnetic viscosity for the MRI when it is active. While the volume occupied by the
field is restricted, the MRI analysis suggests that strong, localized fields may exist at the 
boundary of the iron core at the point of collapse. These fields might play a role in the
collapse process.  In practice, these ST and MRI mechanisms may apply in different 
geometric locations in a given realistic 3D model at a given time, and may leave behind 
fossil magnetic fields in regions that revert from instability to stability. We return to these 
points in \S \ref{discuss}.

Figure \ref{15rot_evolve} presents the MRI diffusion coefficient and the angular velocity 
at the end of hydrogen burning, helium burning, oxygen burning, silicon burning, and at 
the onset of core collapse for the 15 \ms\ model with only the magnetic effects of the MRI. 
At the end of core helium burning, locations A, B, and C in panel two on the right denote 
regions where shear triggers the MRI. Sufficiently steep gradients in $\Omega$ can overcome 
strong composition buoyancy stability, but shallower gradients in $\Omega$ suffice where 
the composition gradient is less steep, specifically in regions where a lighter composition 
has nearly merged into a heavier one.

Figure \ref{15terms_He} shows the distribution of the three components that contribute to the
instability criterion for the MRI from Equation (\ref{MRIcrit}) at the end of core helium burning
for the 15 \ms\ model with only the magnetic effects of the MRI. The terms are $(\eta/\kappa)N_T^2$ (black), 
$N_{\mu}^2$ (red), and $2 q \Omega^2$ (green). The first two terms are stabilizing terms 
(except in convective regions where the first term is a driving term); the third term is the driving 
term for the MRI. Because the condition of MRI instability is so sensitive to gradients, the results 
are sensitive to the finite differencing associated with zoning. To mute this artificial 
effect, we have binned the values of the three terms in Figure \ref{15terms_He} with 
a running top--hat average over 10 zones. The result shows that while the stability is 
sensitive to zone by zone variation, the overall effect is reasonably robust.

Figure \ref{15terms_He} shows that the first, thermal buoyancy term is essentially negligible throughout 
the structure at the phase illustrated since the coefficient $(\eta/\kappa)$ is so small. The competition to 
drive the MRI is between the composition buoyancy stabilizing term and the shear driving term. 
Comparing locations A, B, and C in Figures \ref{15rot_evolve} and \ref{15terms_He} shows
the sensitivity of the MRI to local conditions. Region A from about 4 to 5 \ms\ is all
unstable. The strong composition gradient at 4.05 \ms\ is still not quite enough to
stabilize the structure. Region B has only a mild shear, but the composition gradient is 
correspondingly weaker and this whole extended region from about 2 to 4 \ms\ is unstable; 
the shear term dominates the buoyancy term throughout region B. Region C corresponds to the innermost 
small steep rise in $\Omega$ in Figure \ref{15rot_evolve}. Despite the increase in shear, 
inspection of Figure \ref{15terms_He} shows that the buoyancy dominates there and the small 
region right at a mass of 2.03 \ms\ is stable, but that the structure is unstable on 
both sides of that spike in structure. Interior to 1.1 \ms, the shear is very small 
and the structure is stable. 

Figure \ref{15j} shows the final respective distributions of angular momentum per 
unit mass for the 15 \ms\ models with no rotation, rotation but with the magnetics effects 
suppressed, with ST only, with MRI only, and with both ST and MRI implemented. 
 The top two panels show that for these cases there is very little change in the
angular momentum distribution after oxygen burning; the lines for post--oxygen
burning, post--silicon burning and the final model are basically indistinguishable.
The models with MRI only and ST plus MRI show that there is some evolution from
oxygen burning to silicon burning to core collapse, specifically induced by the MRI.

Figure \ref{15MRIall} shows the distribution just prior to core collapse of the 
composition, the angular velocity, the diffusion coefficients, and the components 
of the MRI instability criterion for the 15 \ms\ model with the MRI, but not ST, active. 
The upper left panel shows composition (from H to Fe), the upper right panel shows
the profile of the angular velocity, $\Omega$, the lower left panel shows the 
logarithm of the diffusion coefficients (for mixing) for the various processes and 
the lower right panel shows a comparison of the three terms of the radial MRI instability
criterion of Equation (\ref{MRIcrit}). Note that the very center is iron rich. This model has 
proceeded up to the brink of iron--core collapse. Figure \ref{15STall} gives the same 
distributions for the model with the ST, but not MRI, active and Figure \ref{15MRISTall} 
when both the MRI and ST are active. The model with ST only has smaller angular velocity 
in the center than the model with MRI only and essentially negligible rotation beyond 
that. The model with both MRI and ST active has a very similar final angular profile,
but there are quantitative differences in all the distributions. 

The rapid jumps by orders of magnitude in the diffusion coefficients and in the 
thermal buoyancy, $N_T$, seen in the models are ``real" and caused by rapid change 
in the shear and the composition at boundaries. There is a question as to whether or
not these features are adequately resolved in our calculations. We have done some 
resolution studies in the 15 \ms\ model by altering the parameter 
{\bf delta{\textunderscore}mesh{\textunderscore}coeff} in MESA that 
controls the spatial zoning resolution. The original value was 0.5. We both increased and 
decreased the resolution, with values of 0.25 and 0.7 and found no perceptible difference 
in the resulting angular velocity profiles at the onset of core collapse. We then tried a value 
of 0.1, both with and without our MRI prescriptions. At such high resolution, about 30,000 
zones, the code crashed before even getting through core helium burning. The computation 
of derivatives becomes unstable. Future studies should investigate these jumps in the
diffusion coefficients more carefully at higher resolution, perhaps by isolating the regions 
of strong gradients in a dedicated simulation rather than attempting a ``whole star" approach 
as we have done here. The true physical structure is surely multidimensional, requiring 
appropriately higher resolution to resolve. We note that while this issue arises in the context 
of the MRI, it also probably pertains to ST and other magnetic effects that are inherently 
multidimensional and worthy of more careful study.

\subsection{20 \ms\ Model}

The models corresponding to ZAMs mass of 20 \ms\ also proceeded up to the brink of iron--core collapse.
Figure \ref{20j} shows the final respective distributions of angular momentum per unit mass.  
As for Figure \ref{15j}, the MRI alone or in tandem with the ST process affects the evolution
of the angular momentum distribution from oxygen burning to silicon burning to the final onset
of collapse in a way that ST alone does not. Note in the lower right panel that with both the ST
and MRI active, there is a substantial increase in the angular momentum in the vicinity of what 
had been the outer edge of the helium core at around 6 \ms. As illustrated below, this is because the
combined effect of the two mechanisms homogenizes the outer structure.

Figure \ref{20rotation} gives the distributions of angular velocity, $\Omega$, and the equatorial 
velocity for the models with ZAMS mass of 20 \ms\ with rotation but with the magnetic effects 
suppressed, with ST but not MRI, with MRI but not ST, and with both ST and MRI implemented. 
The MRI alone can result in considerable spin--down of the inner core compared to a rotating 
model with no magnetic effects, in contrast to the models for MRI only in the 7 and 11 \ms\
models. The dash--dotted lines correspond to the case where both ST and MRI are active.
The angular momentum per unit mass is constant beyond $\sim 3$ \ms, a consequence of
the mixing of the helium core and the outer envelope.

Figure \ref{20MRIall} shows the distribution just prior to core collapse of the 20 \ms\ models of the 
composition, the angular velocity, the diffusion coefficients, and the components of the MRI instability 
criterion for the model with the MRI, but not ST active. Figure \ref{20STall} gives the same distributions 
for the model with the ST, but not MRI, active and Figure \ref{20MRISTall} when both the MRI and 
ST are active. The iron core is of about the same mass in all three magnetic models, but the 
oxygen core is somewhat larger in the model with ST only, $\sim 3.8$ \ms\ versus $\sim 
3.2$ \ms\ for the other two models. In Figures \ref{20MRIall}, \ref{20STall}, and \ref{20MRISTall}, 
the center of the iron core spins slightly slower for the model with the MRI only than for that with 
ST only, but slower yet for the model with both magnetic effects. In these final models, the MRI is 
not active in the inner core, as may be seen by inspection of the lower panels of the figures that 
give the diffusion coefficients and the contributions to the MRI.

The helium core is about 6 \ms\ for both the models with MRI only and ST only, but for the 
model with both effects, the H/He envelope extends down to the oxygen--rich layers at about 
3 \ms. With both mechanisms active, the helium shell has been mixed entirely out into the envelope. 
This is consisten with the anomolous distributions of $j$ and $\Omega$ noted in Figures \ref{20j} 
and \ref{20rotation}.  The envelope of this mixed model has a helium abundance of $\sim 50$ \% by 
mass. As a result of the helium enrichment, the model has become a yellow supergiant with a radius 
of $1.1\times10^{13}$ cm and an effective temperature of 7900 K at the point of collapse. Because 
the envelope of this model has contracted, it is also radiative. This can be seen in the lower left panel 
of Figure \ref{20MRISTall}, where the convective region ends at about 11 \ms. Beyond that, the 
radiative envelope is mostly dominated by ES mixing, but the model yields narrow regions where 
the MRI dominates. 

That we only see this complete homogenization of $j$ and composition in the 20 \ms\ model is 
probably because this more massive model is more dominated by radiation pressure, bringing it 
closer to the condition of neutral stability and hence more prone to mixing. It would not be wise 
to take this result too literally, but it suggests that more massive stars would be even more susceptible 
to such homogenization, and that enhanced mixing could yield a population of yellow or even 
blue supergiant supernova progenitors with helium--rich envelopes. Possible implications for 
the paucity of SN~IIP at M $\gta$ 17 \ms\  \citep{smartt09} and for SN~1987A have not escaped us. 

\section{Discussion and Conclusions}
\label{discuss}

We have used the MESA stellar evolution code to compute rotating stellar models with 
magnetic effects due to the Spruit--Taylor mechanism and the MRI, separately and 
together, in a sample of massive star models. We find that the MRI can be active in the post--main
sequence stages of massive star evolution, slowing core rotation and leading to mixing effects that 
are not captured in models that neglect the MRI. The MRI tends not to be active in the cores of
the models at the onset of core collapse, but the structure of those cores can be affected by
the activity of the MRI in previous stages of the evolution.

We find that the MRI is activated throughout the intermediate stages of the evolution of massive stars
as regions arise where there is sufficient shear to overwhelm the stabilizing effects of buoyancy stability. 
The shear tends to be strongest at composition boundaries where the stabilizing effects are also strong.
There are also extended regions where both the buoyancy and the shear are mild, but, nevertheless,
the shear is sufficient to enable the MRI. The issue of when and where the MRI is triggered
is thus a subtle quantitative one. The activity of the MRI may depend rather sensitively on issues 
such as the convective instability criterion, semi--convection, and overshoot. Once the instability
sets in, its effects can spread more broadly beyond the regions of immediate instability, 
leaving changes in the density, temperature, and composition structure.

The MRI acting alone can slow the rotation of the inner core in general agreement with the
observed ``initial" rotation rates of pulsars. In our models, when the ST and MRI mechanisms
are both invoked, the final rotation more closely resembles models with ST alone than
with MRI alone. The dominance of ST over MRI when they are both active is presumably
due to ST being active over larger spatial extent and being less intermittent than MRI. 
This issue is worth more careful future study. The MRI can also serve as an effective mechanism 
for the mixing of different composition layers. Plots of the mean molecular weight, $\mu$, show 
that models with the MRI or with both MRI and ST active tend to produce smoother composition 
profiles in the inner core than ST acting alone.

The magnetorotational effects can move a model from the regime of degenerate C/O cores 
to the regime of degenerate cores of O/Ne/Mg, and hence shift the final evolution from 
thermonuclear explosion to core collapse by electron capture instability. Similar statements 
apply to models that form O/Ne/Mg cores in standard non--rotating, non--magnetic evolution. 
Magnetorotational effects can move a model from the regime of degenerate O/Ne/Mg to the 
iron--core regime. This is especially interesting because work identifying progenitors shows 
that the progenitors of SN~II arise from rather low mass stars $\gta$ 8 \ms. Magnetic effects 
may thus shift the fundamental physics of core collapse in low--mass models.  We have only 
touched on this topic in this exporatory work that sought to establish the proof--of--principle. 
This subject clearly merits deeper study.

\begin{figure}
\begin{center}
\includegraphics[angle=-90,width=16cm]{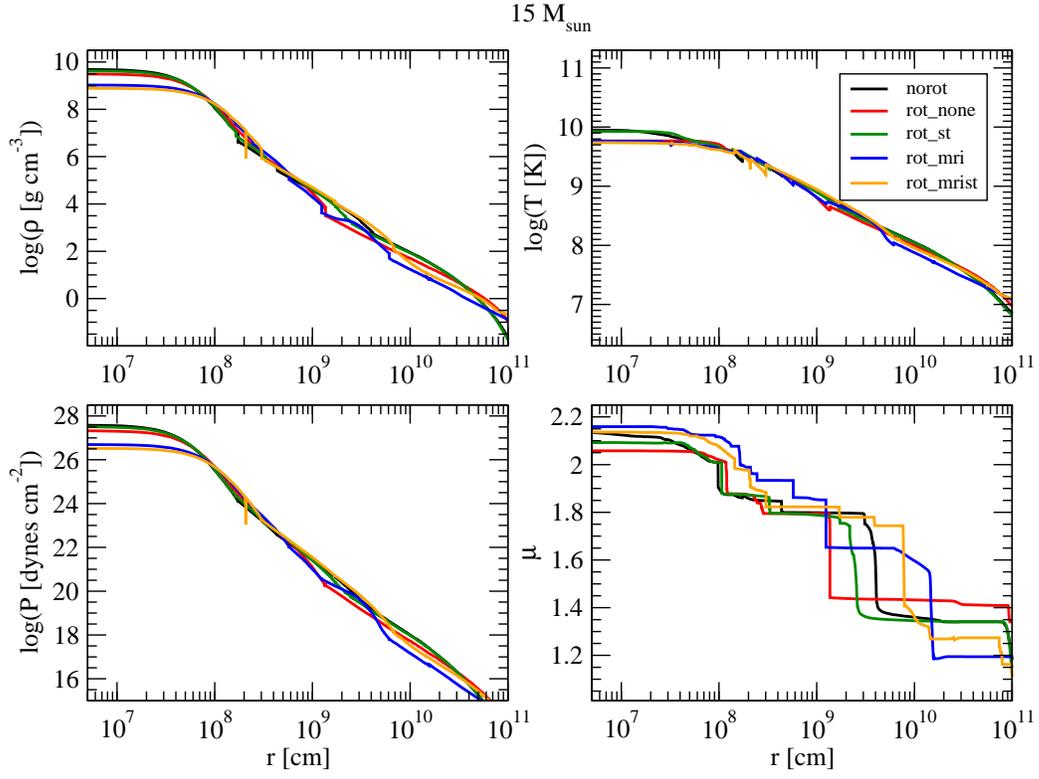}
\caption{Radial distributions of density (upper left), temperature (upper right), pressure 
(lower left) and mean molecular weight (lower right) for the final model corresponding to the 
fiducial model with ZAMS mass of 15 \ms\ for the cases with no rotation, rotation 
but no magnetic effects, ST but not MRI, MRI but not ST, and with both ST and MRI active. See the online version for color.}
\label{15structure} 
\end{center}
\end{figure}

\begin{figure}
\begin{center}
\includegraphics[angle=-90,width=16cm]{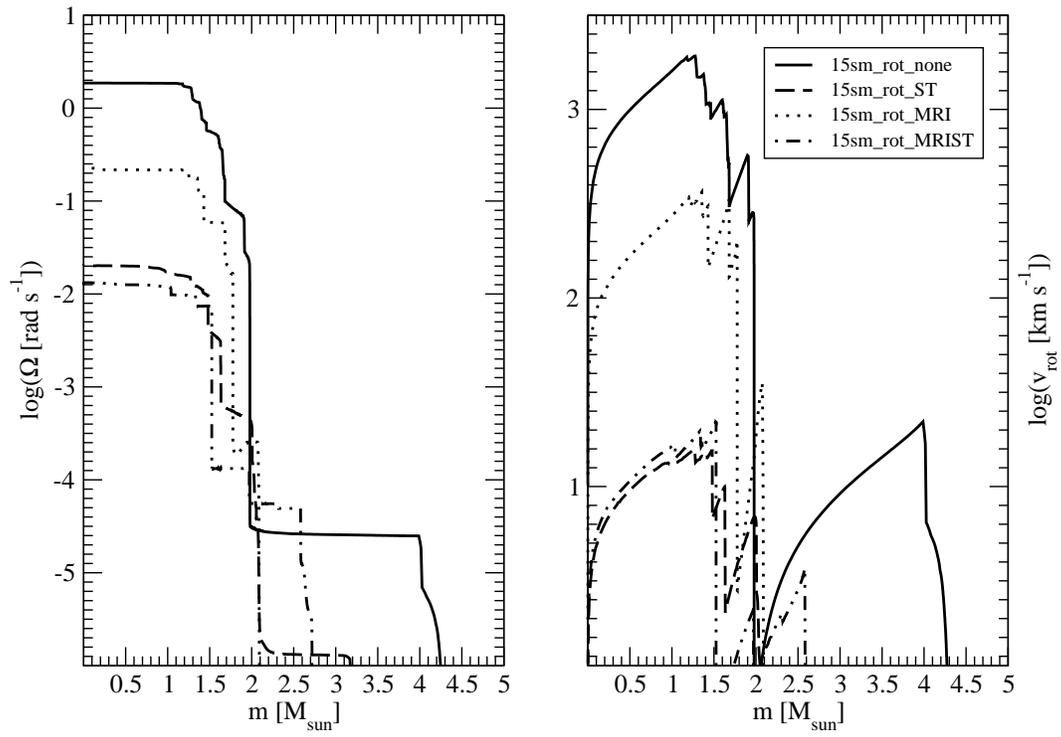}
\caption{Distributions of angular velocity (left) and rotation velocity on the equator (right) at the end of the
calculation of the fiducial rotating model of ZAMS mass of 15 \ms\ for the cases with no magnetic effects (solid
line), ST only (dashed line), MRI only (dotted line), and with both ST and MRI active (dot--dash line). }
\label{15rotation} 
\end{center}
\end{figure}

\begin{figure}
\begin{center}

\includegraphics[angle=-90,width=16cm]{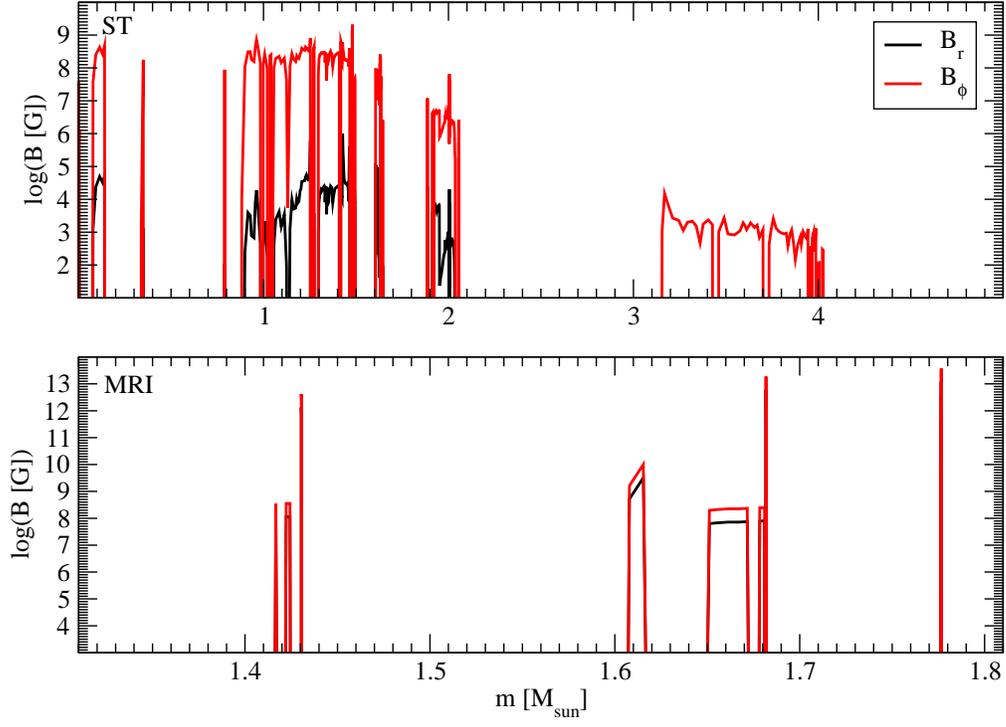}
\caption{Distribution with respect to mass of the magnetic field at the end of the calculation, 
just prior to core collapse, of the fiducial rotating model of 15 \ms. Top, the case where the 
ST, but not MRI, is active; bottom, the case where the MRI, but not ST, is active. Both the toroidal (red) 
and radial (black) fields are presented. For the ST mechanism, the radial field is much less than
the toroidal field, leading to a relatively small magnetic viscosity. For the MRI, the radial field 
is a constant fraction, 0.32, of the toroidal field. Note that while sparsely distributed, the strength of 
the toroidal field in the case of the MRI far exceeds that of the ST mechanism. The relatively large 
radial field contributes to a larger magnetic viscosity in the case of the MRI. See the online version for color.}
\label{15B} 
\end{center}
\end{figure}

\begin{figure}
\begin{center}
\includegraphics[angle=-90,width=16cm]{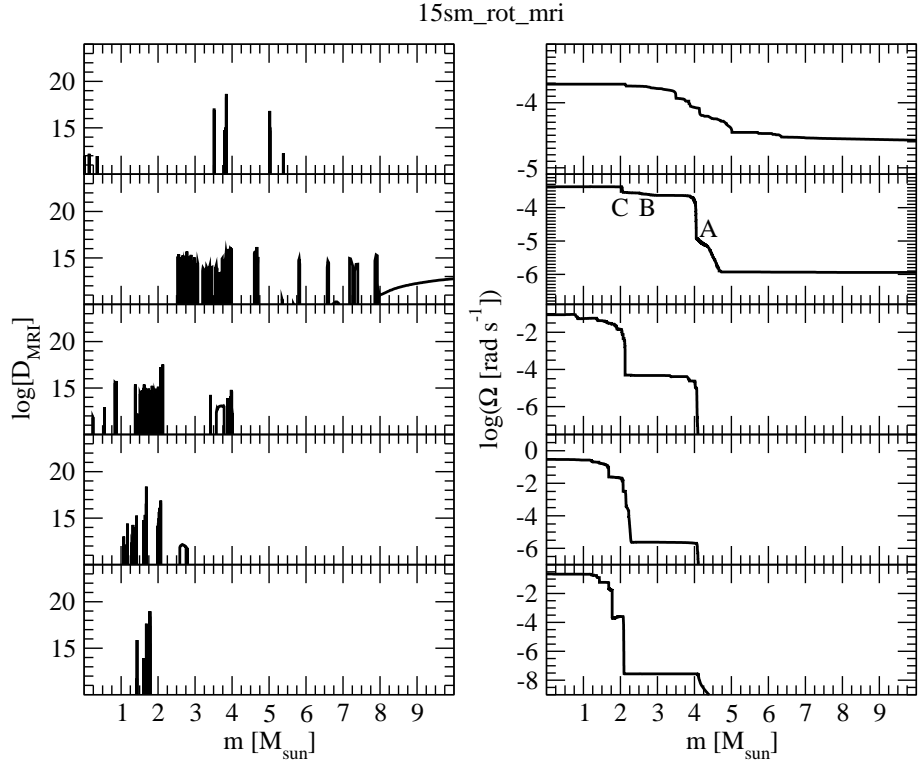}
\caption{Distribution with respect to mass of the MRI diffusion coefficient (left) and angular velocity (right)
at the end of hydrogen burning, helium burning, oxygen burning, silicon burning, and at the onset of core 
collapse (top to bottom) for the fiducial model of ZAMS mass 15 \ms\ for the rotating model with the MRI, 
but not ST, active. For panel 2 on the right, locations A, B, and C denote regions of steep or shallow gradients 
in $\Omega$ where shear generates MRI at the end of core helium burning (see Figure \ref{15terms_He}).}
\label{15rot_evolve} 
\end{center}
\end{figure}

\begin{figure}
\begin{center}
\includegraphics[angle=-90,width=16cm]{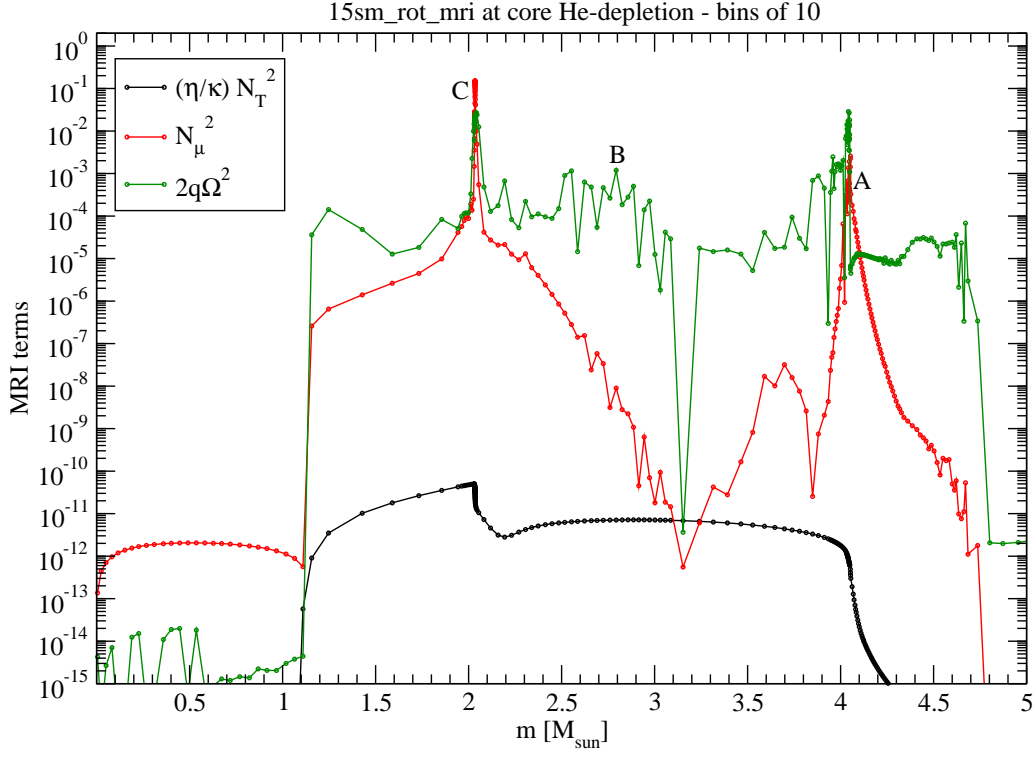}
\caption{Distribution with respect to mass of the terms that contribute to the instability
criterion of the MRI at the end of core helium burning of the fiducial rotating model of ZAMS mass
15 \ms\ for the case where the MRI, but not ST, is active. The terms are $(\eta/\kappa)N_T^2$ (black), 
$N_{\mu}^2$ (red), and $2 q \Omega^2$ (green), where the latter is the driving term for the MRI.
Locations A and B denote regions of steep or shallow gradients in $\Omega$ where shear generates 
MRI (see Figure \ref{15rot_evolve}). The small step in $\Omega$ and hence shear at point C at 2.03 
\ms\ in Figure \ref{15rot_evolve} is stabilized by the composition buoyancy there, but regions on 
either side are unstable. See the online version for color.}
\label{15terms_He} 
\end{center}
\end{figure}

\begin{figure}
\begin{center}
\includegraphics[angle=-90,width=16cm]{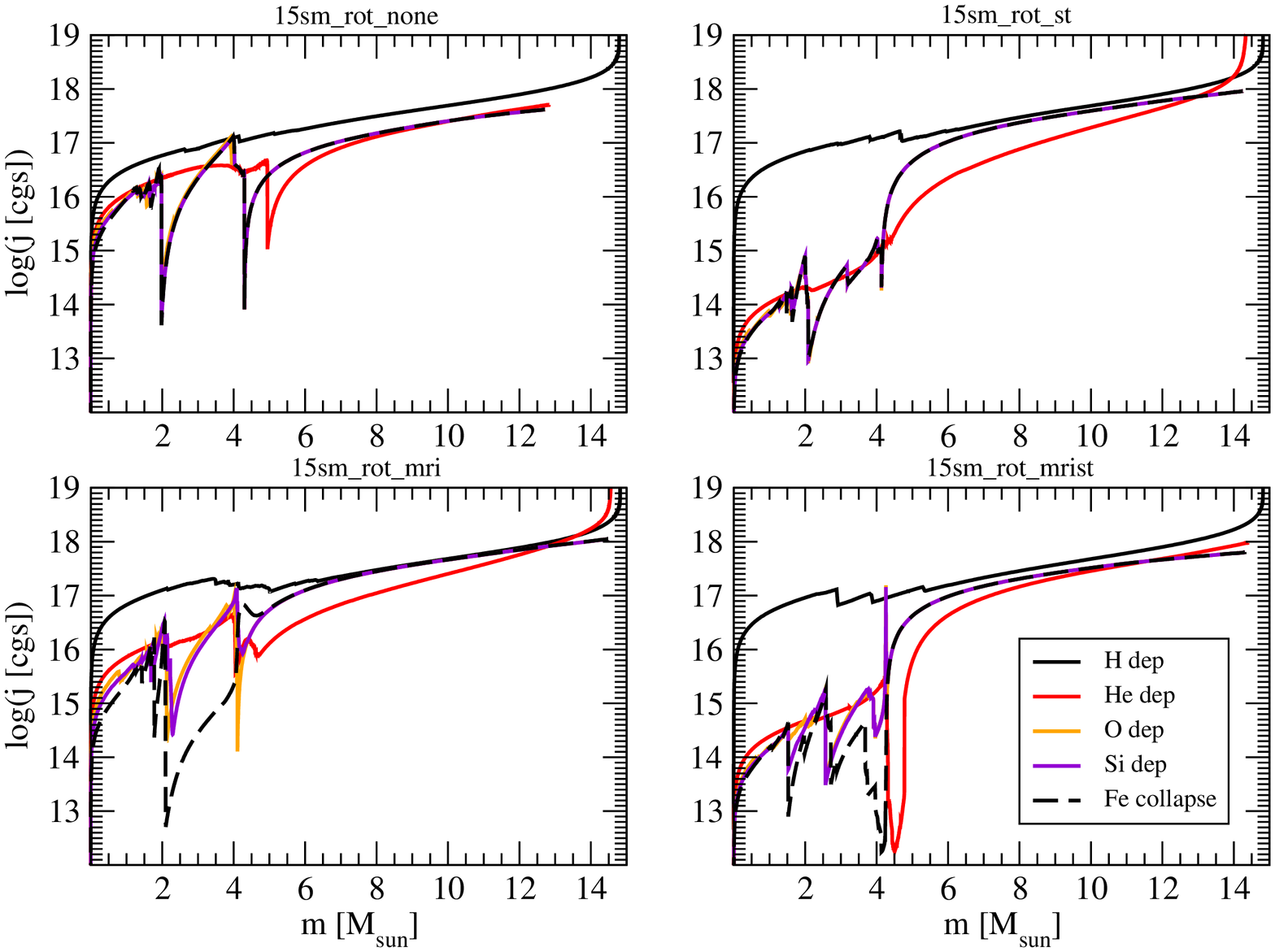}
\caption{Distribution with respect to mass in the fiducial model with ZAMS mass of 15 \ms\ of the specific angular momentum
at the end of hydrogen burning, helium burning, oxygen burning, silicon burning, and at the onset of core collapse
for the cases with rotation but no magnetic effects (upper left), ST but not MRI (upper right), MRI but not ST (lower left), 
and with both ST and MRI active (lower right). See the online version for color.}
\label{15j} 
\end{center}
\end{figure}

\begin{figure}
\begin{center}
\includegraphics[angle=-90,width=16cm]{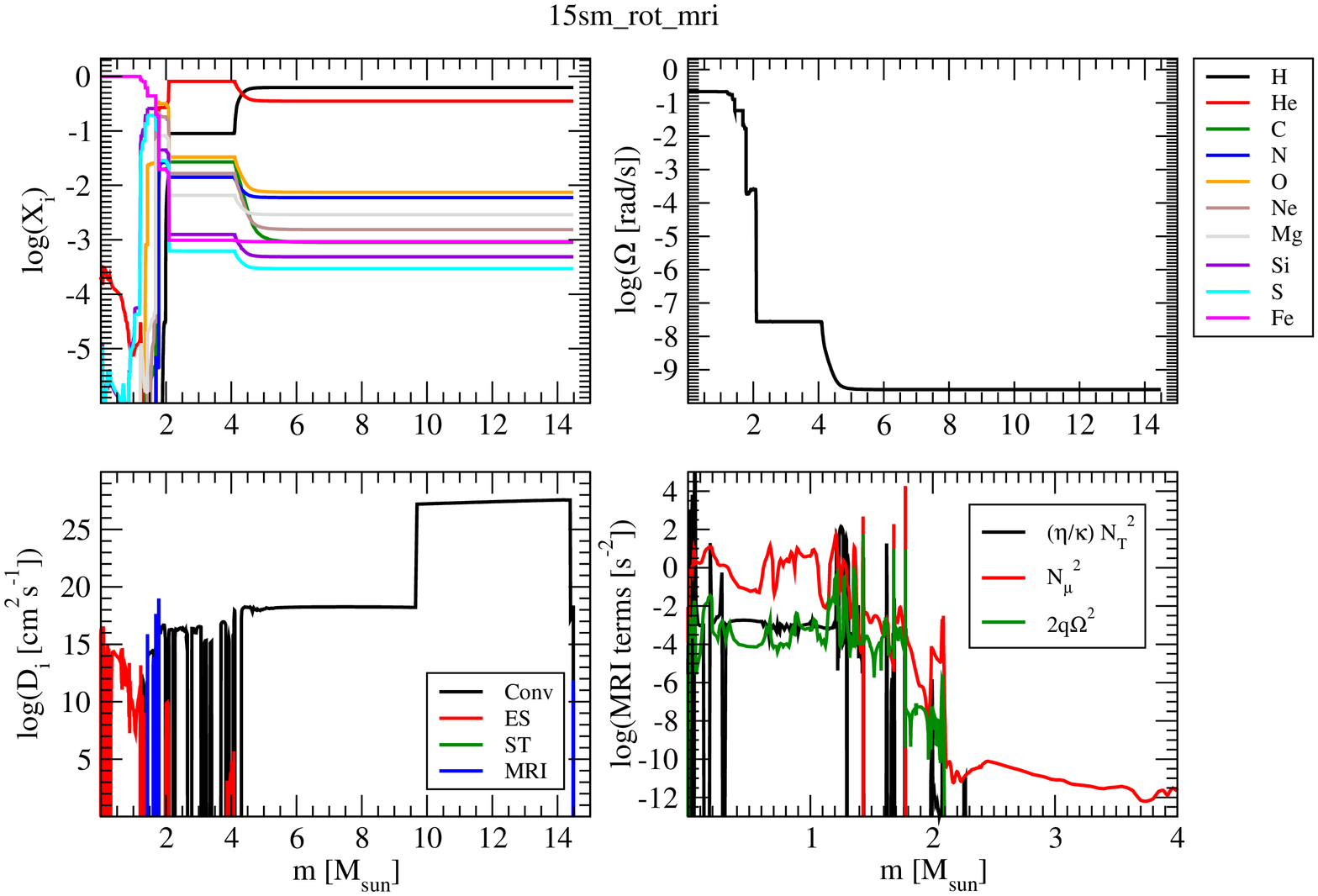}
\caption{Distribution with respect to mass in the model with ZAMS mass of 15 \ms\ of the final
distributions of composition (upper left), angular velocity (upper right), diffusion coefficients 
(lower left) and the components of the MRI instability criterion (lower right) for the model with 
the MRI, but not ST, active. See the online version for color. }
\label{15MRIall}
\end{center}
\end{figure}

\begin{figure}
\begin{center}
\includegraphics[angle=-90,width=16cm]{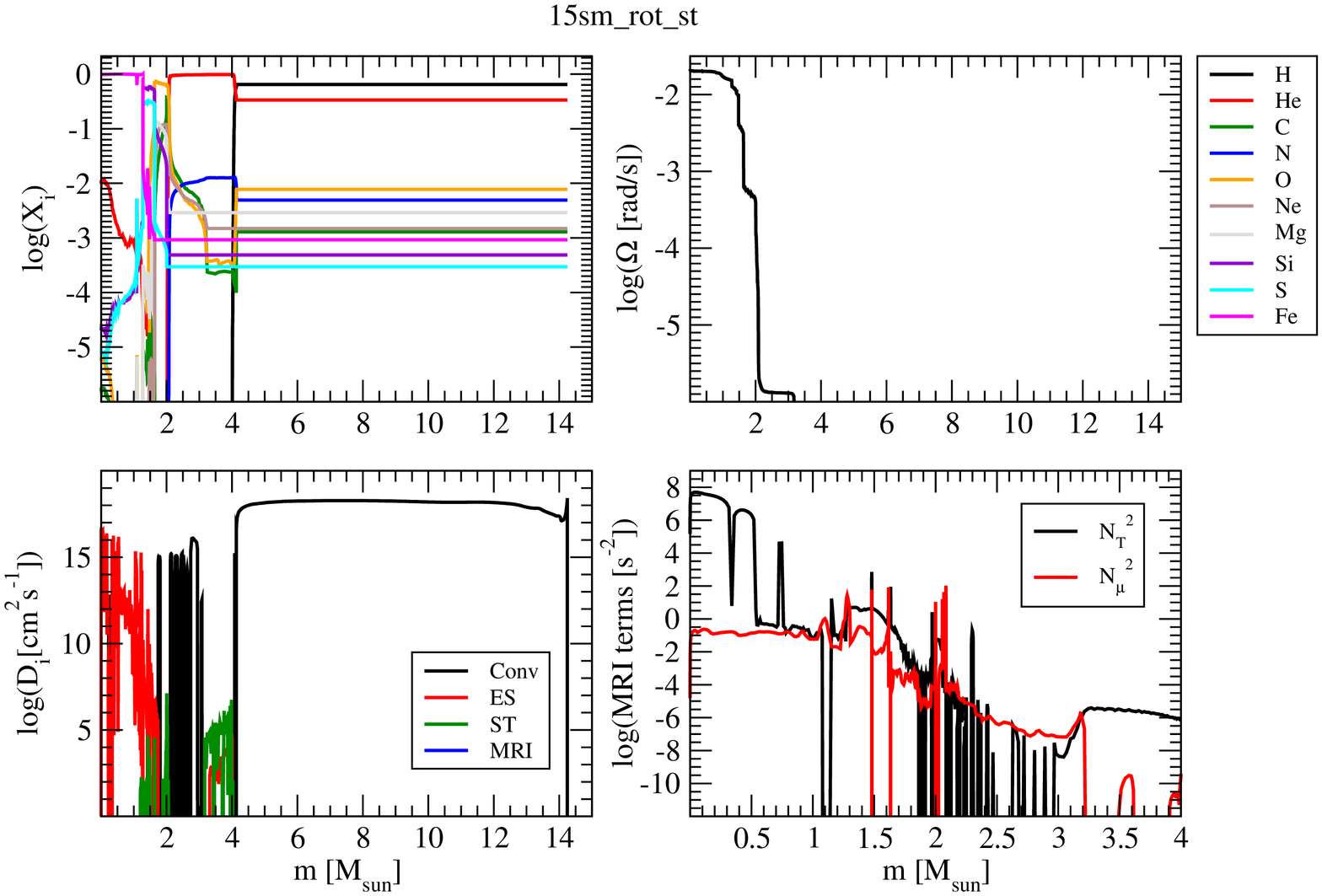}
\caption{Distribution with respect to mass in the model with ZAMS mass of 15 \ms\ of the final
distributions of composition (upper left), angular velocity (upper right), diffusion coefficients 
(lower left) and the thermal and composition components of the Brunt--V\"{a}is\"{a}l\"{a} frequency 
(lower right) for the model with the ST, but not MRI, active. See the online version for color.}
\label{15STall} 
\end{center}
\end{figure}

\begin{figure}
\begin{center}
\includegraphics[angle=-90,width=16cm]{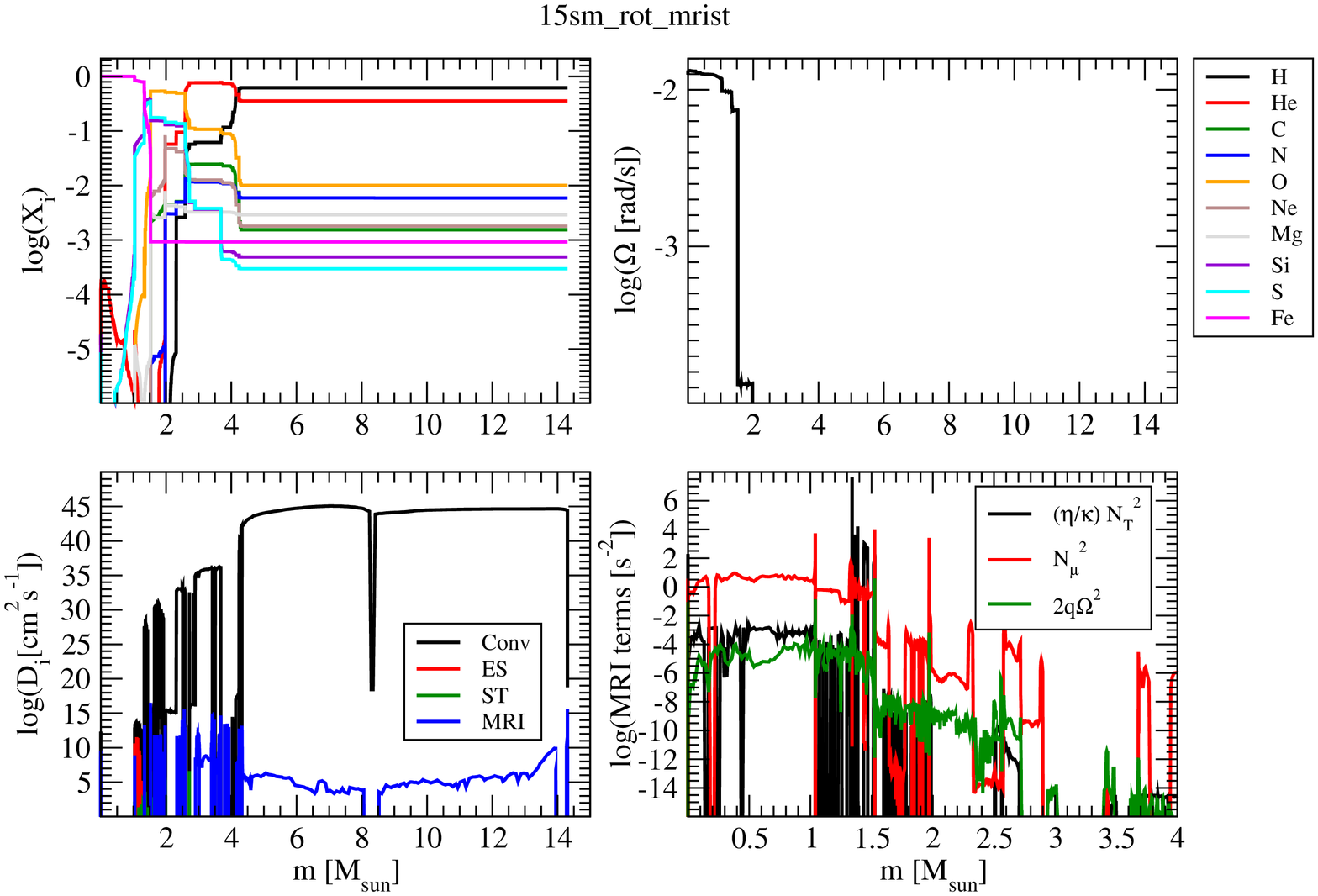}
\caption{Distribution with respect to mass in the model with ZAMS mass of 15 \ms\ of the final
distributions of composition (upper left), angular velocity (upper right), diffusion coefficients 
(lower left) and the components of the MRI instability criterion (lower right) for the model with 
both ST and MRI active. See the online version for color. }
\label{15MRISTall} 
\end{center}
\end{figure}

\begin{figure}
\begin{center}
\includegraphics[angle=-90,width=16cm]{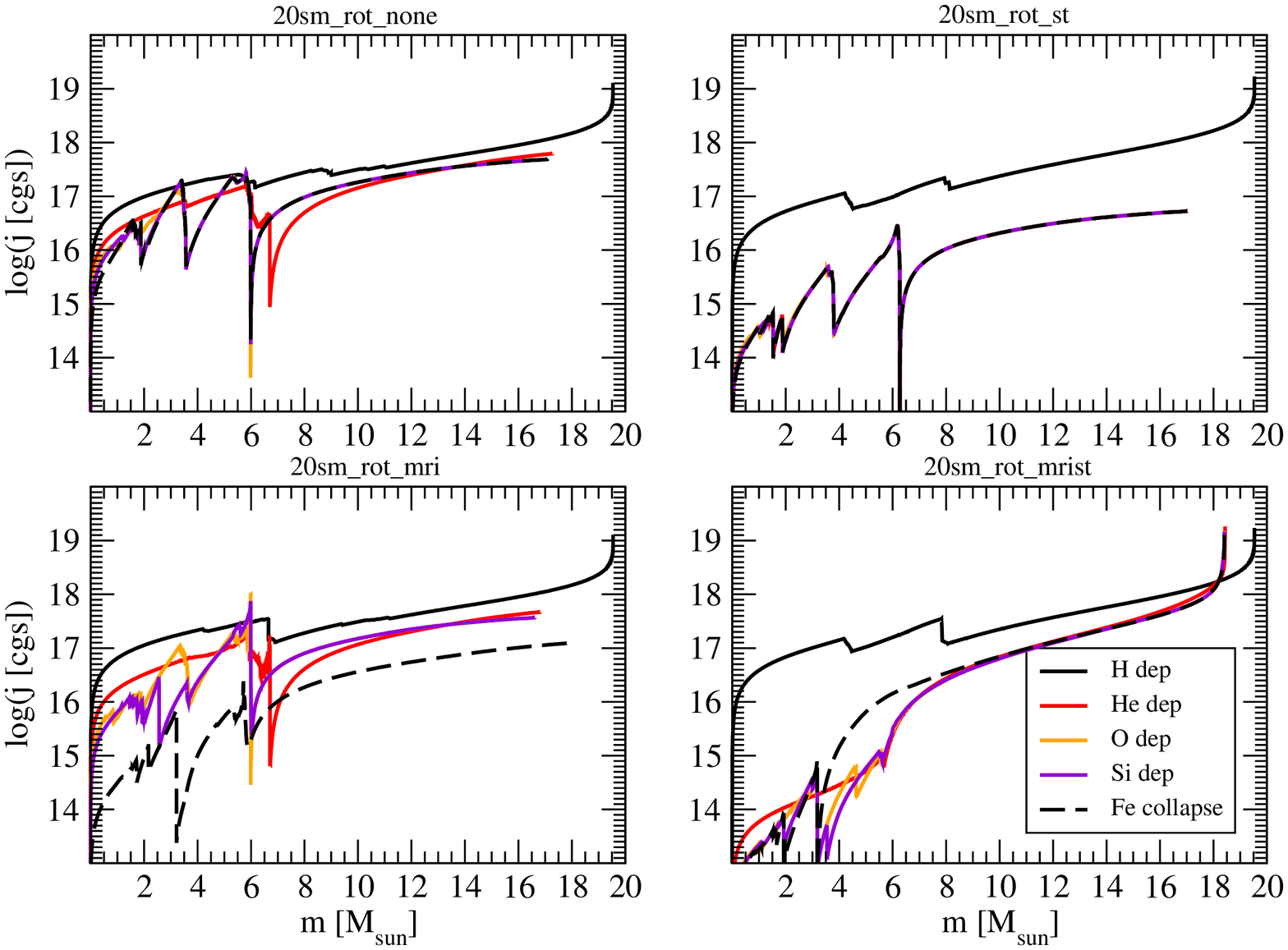}
\caption{Distribution with respect to mass in the model with ZAMS mass of 20 \ms\ of the 
angular momentum per unit mass at the end of hydrogen burning, helium burning, oxygen 
burning, silicon burning, and at the onset of core collapse for the cases with rotation but no 
magnetic effects (upper left), ST but not MRI (upper right), MRI but not ST (lower left), and 
with both ST and MRI active (lower right). See the online version for color. }
\label{20j} 
\end{center}
\end{figure}

\begin{figure}
\begin{center}
\includegraphics[angle=-90,width=16cm]{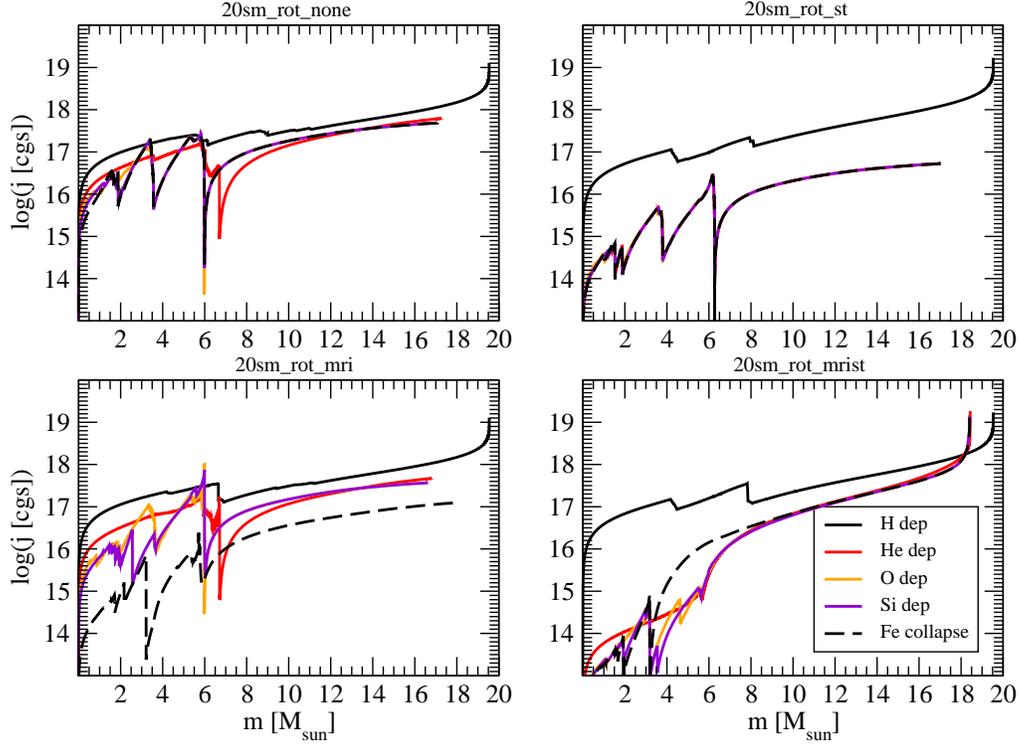}
\caption{Distributions of angular velocity (left) and rotation velocity on the equator (right)
at the end of the calculation of the rotating model of 20 \ms\ for the cases 
with no magnetic effects (solid line), ST but no MRI (dashed line), MRI but no ST (dotted line), 
and with both ST and MRI active (dot--dash line). }
\label{20rotation} 
\end{center}
\end{figure}

\begin{figure}
\begin{center}
\includegraphics[angle=-90,width=16cm]{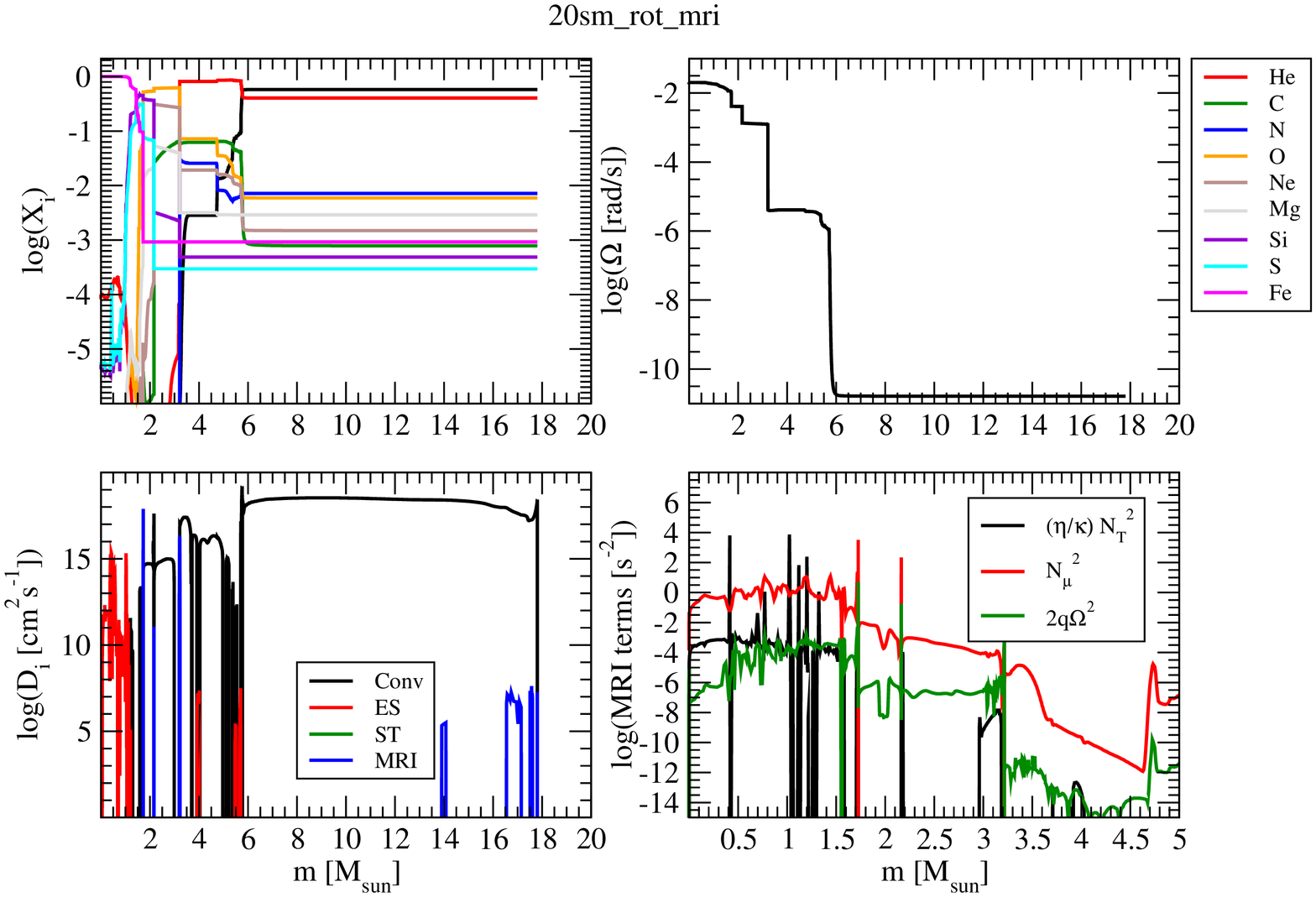}
\caption{Distribution with respect to mass in the model with ZAMS mass of 20 \ms\ of the final
distributions of composition (upper left), angular velocity (upper right), diffusion coefficients 
(lower right) and the components of the MRI instability criterion (lower right) for the model with 
the MRI, but not ST, active. See the online version for color. }
\label{20MRIall} 
\end{center}
\end{figure}

\begin{figure}
\begin{center}
\includegraphics[angle=-90,width=16cm]{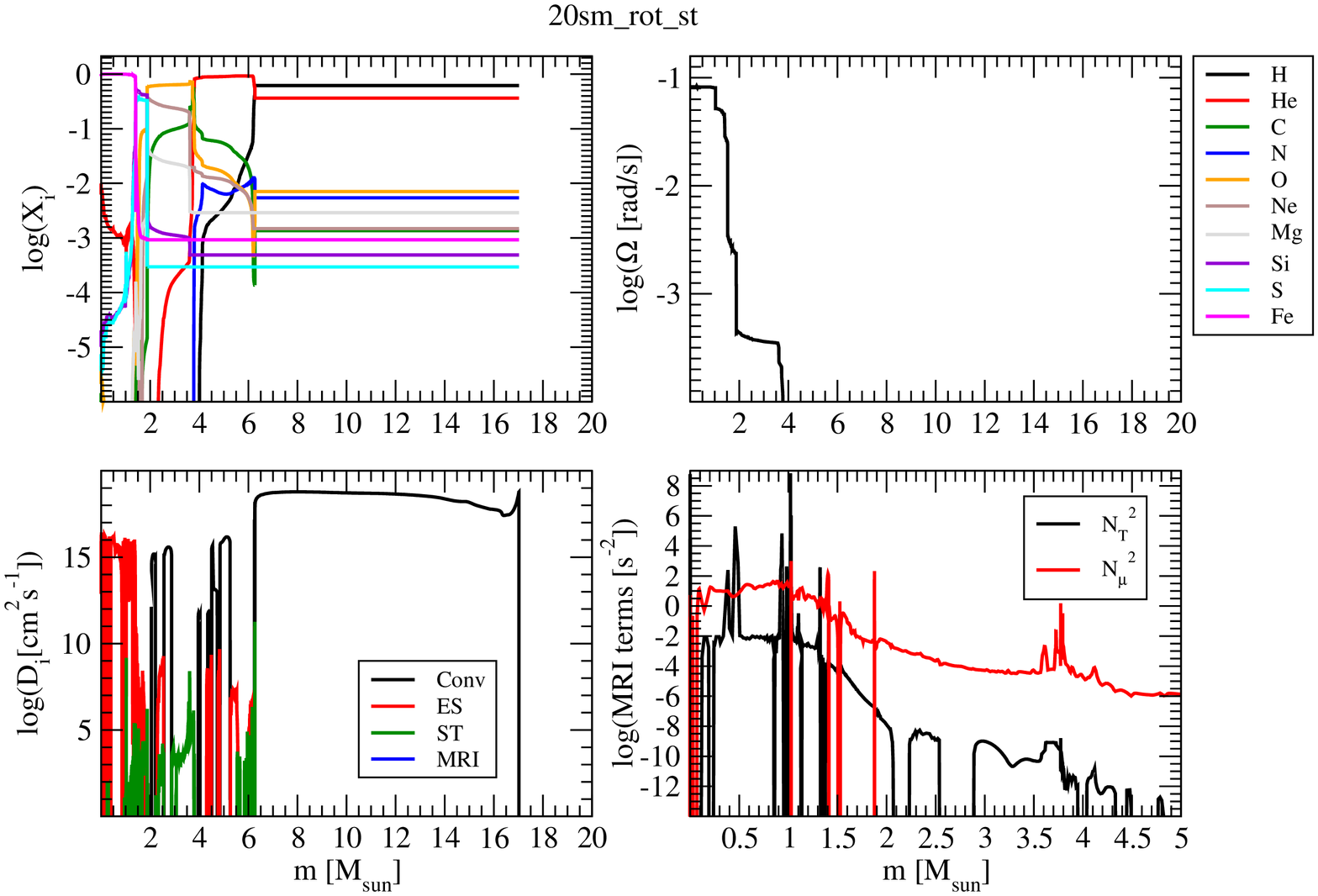}
\caption{Distribution with respect to mass in the model with ZAMS mass of 20 \ms\ of the final
distributions of composition (upper left), angular velocity (upper right), diffusion coefficients 
(lower right) and the thermal and composition components of the Brunt--V\"{a}is\"{a}l\"{a} frequency 
(lower right) for the model with ST, but not MRI, active. See the online version for color. }
\label{20STall} 
\end{center}
\end{figure}

\begin{figure}
\begin{center}
\includegraphics[angle=-90,width=16cm]{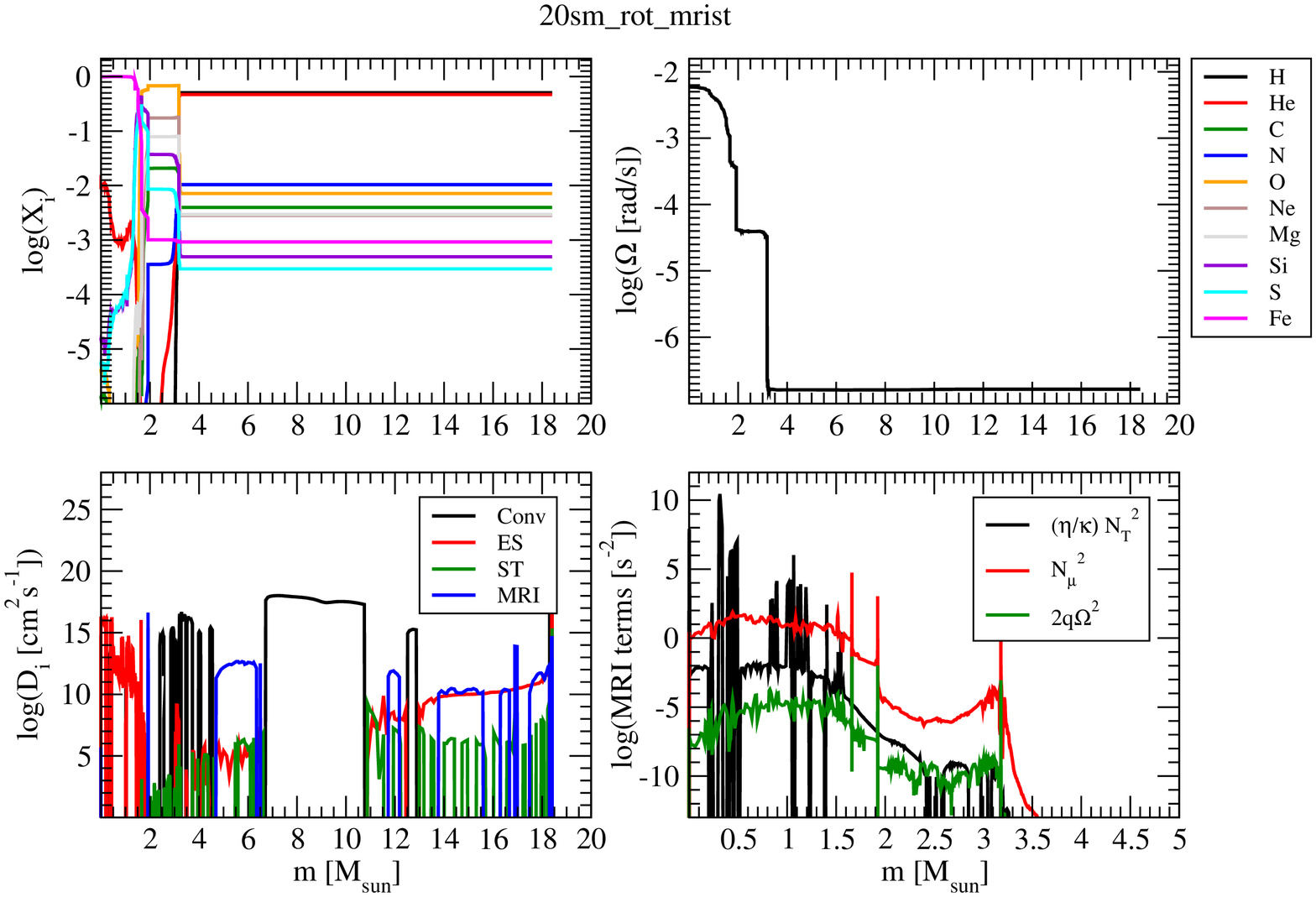}
\caption{Distribution with respect to mass in the model with ZAMS mass of 20 \ms\ of the final
distributions of composition (upper left), angular velocity (upper right), diffusion coefficients 
(lower right) and the components of the MRI instability criterion (lower right)
for the model with both ST and MRI active. See the online version for color. }
\label{20MRISTall} 
\end{center}
\end{figure}

There is a growing understanding that models may be more easy to explode if they are more
``compact," that is, when the density gradient is larger at the edge of the core \citep{OO11,ugli12}.
There are suggestions here that the MRI leads to more compact structure (Figure \ref{15structure}). 
The likelihood that burning proceeds on a convective time scale leading to intermittent, chaotic 
burning \citep{AM11,CO13} may also affect field generation in the late stages. These are both 
topics worthy of deeper study. 

As convective cores contract and begin to spin up and rotate more rapidly than outer radiative layers,
the MRI will come into play, growing seed fields exponentially rapidly to MRI saturation limits
consistent with the thermal and composition gradients that contribute to the local Brunt-V\"{a}is\"{a}l\"{a} 
frequency. Our results suggest that the MRI could already play some role during hydrogen burning
and becomes broadly active by the end of core helium burning. If the MRI provides the effective torque 
and effective viscosity that we estimate, then angular momentum will be advected outward, leading 
to more slowly--rotating, but magnetized, inner cores.

There are many magnetorotational issues in stellar evolution, the proper exploration of which remains 
beyond the state of the art. As \citet{spruit02} emphasized, magnetic instabilities are characteristically 
strongly anisotropic. It is an important first step to include magnetic viscosity effects in spherical 
``shellular" calculations as done in other work and as we do here, but the physics of these instabilities 
ultimately requires investigation in full three--dimensional MHD simulations. 

A variety of issues remain open in the analysis of the ST mechanism itself. \citet{MM05} noted that 
it is very difficult to understand how the ST instability interacts with meridional circulation. \citet{DP07} 
again explored the assumptions and formulation of the ST mechanism. They examined the basic heuristic 
assumptions in the model and questioned whether the dispersion relation can be extrapolated to horizontal 
length scales of order the radius of the star. They presented transport coefficients for chemical mixing 
and angular momentum redistribution by magnetic torques that were significantly different from previous 
published values. Their magnetic viscosity was 2-3 orders of magnitude smaller than that derived by 
\citet{spruit02}. They found the magnetic angular momentum transport by this mechanism to be sensitive 
to gradients in the mean molecular weight. They note that solar models including only this mechanism 
possess a rapidly rotating core, in contradiction with helioseismic data. They conclude that the ST 
mechanism may be important for envelope angular momentum transport, but that some other process 
must be responsible for efficient spin-down of stellar cores. More recently, \citet{Ca14} have
noted that asteroseismology based on {\sl Kepler} observations suggests that the internal rotation
rates of solar--type stars are too low to match the predictions of current rotating models, including the ST
mechanism. The MRI is one candidate to contribute to this extra dissipation.

Another issue is that the predicted field 
structure for the ST mechanism has a radial field that is weaker than the toroidal field by a factor of order 
$10^4$. While one expects rotation about an axis and associated shear to produce predominantly 
toroidal field, this extreme ratio of toroidal to radial field is, to the best of our knowledge, unprecedented 
in numerical simulations. As an example, \citet{Braith06} modeled the ST process and found that the dynamo 
worked as predicted, but the resulting radial field (cylindrical or spherical) was of order 20\% of the toroidal 
component (whereas \citet{Za07} found an instability, but no dynamo).  In conditions where the background 
varies sufficiently slowly, the equilibrium field structures found by \citet[][see also Mitchell et al. 2014]{Braith09} 
may also be relevant. In those solutions characterized by a twisted torus and a poloidal component, the radial 
component is again a substantial fraction of the total field. Understanding the radial component of the 
field is important because that is the component that determines the magnetic torque and hence 
the effective magnetic viscosity. Clearly, the effective magnetic viscosity will be substantially larger if B$_r$ 
is a substantial, not a tiny, fraction of B$_{\phi}$.

Related issues plague the proper treatment of the MRI. In our current models, we have used prescriptions 
for the ST and MRI separately and together, but have not attempted to understand the fundamental, perhaps 
non--linear interaction of these instabilities. The ST instability and the MRI may occur in different regions 
of the star, the ST instability near the rotation axis and the poles, the MRI perhaps at lower latitudes. In 
regions where the two mechanisms may both operate, the MRI will be more rapid, but then enhance the 
field to the saturation limit where $\omega_A \sim \Omega$, at which point the stronger field will also 
enhance the effective viscosity of the ST mechanism. We do not capture this sort of interaction in the 
current models. The full interplay of both of these instabilities with convection, semiconvection, 
thermohaline instabilities, radiation pressure other dynamo processes, 
and meriodional circulation in 3D is a complex one that will be a challenge to explore.

We have invoked here the local instability criterion for the MRI (Equation \ref{MRIcrit}), but a proper analysis
of the MRI instability should be a global analysis as outlined by \citet{Pino08}. Global analyses can reveal 
that conditions that appear locally unstable to the MRI are not, in fact, unstable, for instance because 
the unstable wavelength will not fit into the finite radial region of instability. 

Because it is very difficult to resolve the most rapidly--growing modes of the MRI in core collapse, 
many MHD simulations invoke very strong initial fields, $\sim 10^{12}$ G, so that compression and 
wrapping effects mock up the final fields expected from the MRI \citep{burrows07, mosta14}. If the
pre--collapse seed fields are more modest, this is not a proper procedure since the MRI is expected
to grow fields exponentially rapidly on a post--collapse time scale, $\sim \Omega^{-1}$, much
more rapid than the collapse and wrapping timescales. In this context it is interesting to note
that our MRI models lead to fields at the boundary of the iron core of $\sim 10^{12}$ G. These
primarily toroidal fields may exist only in thin layers with a distribution very different than
a dipole. The effect of such fields on magnetic core collapse is clearly of great interst.

The effect of the MRI on on the evolution preceeding core collapse may have implications for a host of 
issues related to neutron star formation, for instance the initial spins of pulsars and the mechanism 
of the formation of magnetars. Our models suggest rather slowly rotating iron cores, which cannot be 
ruled out. This is because of the very interesting possibility raised by \citet{blo03} and \citet{BM07} 
that collapse triggers the standing accretion shock instability, SASI, and that in 3D, the SASI can lead 
to fairly rapidly rotating neutron stars even in cases where the original iron core has very small or no 
angular momentum. If the proto--neutron star is spun up in this way, the MRI may again be triggered  
as discussed by \citet{aki03}, \citet{Ober09}, \citet{SY14} and others. The MRI in concert with field 
compression and wrapping effects could provide the magnetic fields of pulsars. The rotation that can 
be induced by the SASI may not be enough yield a Rossby number (the ratio of convective overturn 
time to rotational period) of order unity and hence a vigorous $\alpha-\Omega$ dynamo as invoked 
by \citet{DT92} to account for magnetar-level fields, but the MRI may be able to do so under more 
modest spin conditions.      

The rotational profile at the time of core collapse is not the only important ingredient in the problem of 
determining the significance of the MRI. If the MRI does play a role in the final evolution of rotating 
stars, it is not sufficient to invoke it at the end of a calculation where steep gradients of angular velocity 
are already built up; it must be applied from the beginning. The magnetic field developed in earlier 
phases may linger even after a given mass layer becomes stable to the MRI (or to ST). If, in the prior 
evolution, there were a portion of the structure that triggered the MRI, the field would rapidly grow
to saturation. If the rotational structure then flattens to small q because of the effective magnetic 
viscosity, there might be a fossil rather large, mostly toroidal, field left behind. The latter might then 
affect the subsequent rotational evolution and the field in the progenitor at the time of collapse. If that 
were the case, then one needs to follow the whole evolution of the star, including fossil MRI regions, 
to know the rotational and magnetic state at the time of collapse. 

A key question is then the time scale for magnetic field dissipation. If the field decays only
through the processes of magnetic diffusivity, then the characteristic timescale can be written,
using Equation (\ref{eta}), as
\begin{equation}
\tau_{diff} \sim \frac{\ell^2}{\eta} \sim \frac{H_p^2}{\eta} \sim 1.3\times10^{10}~ {\rm y}~ H_{p,9}^2T_8^{3/2},
\end{equation}
where $T_8$ is the temperature in units of $10^8$ K. This is a very long time and if this were 
the relevant physics, the fossil fields would be significant. If the field decays through reconnection, 
perhaps a more likely circumstance, then the time scale could be much shorter. The reconnection 
physics under the conditions of interest is not known, but we can make an estimate based on a 
simple model for resisitive reconnection \citep{kulsrud05,bellan06}
\begin{equation}
\label{reconn}
\tau_{reconn} \sim \sqrt{\tau_{diff} \tau_A} \sim 100~{\rm y}~H_{p,9}^{3/2} T_8^{3/4} \rho^{1/4} B_8^{-1/2},
\end{equation}
where $B_8$ is the field strength in units of $10^8$ G. This implies that for the fiducial conditions
chosen in Equation (\ref{reconn}) the timescale could be short and the fossil fields would decay
quickly compared to an evolution time scale over most of the evolution. This may not be the case late in the
evolution when the density is high, depending on the field strength. Fossil fields might be important
in the last several centuries of the life of a massive star, when other complications in the evolution
such as burning on convective time scales are also likely to exist. 

An area of great impact is the quest to understand the role of stellar collapse in the formation of 
cosmic gamma-ray bursts. In particular, the results here suggest that slow rotation is the rule and hence 
that ``collapsar" models \citep{woosley93} that require rather rapid rotation of a newly--formed black 
hole and its associated accretion disk could be problematical. As outlined above, there might be a route 
to form magnetars, with their potential role in the long, soft GRB phenomenon, if the SASI generates original 
neutron star spin. Even this possibility would raise a host of problems since not all collapse leads to 
magnetars and the rate of birth of GRBs is substantially less than that estimated for magnetars. Even if 
one contemplates a magnetar origin for GRBs \citep{mazzali14}, the issue of what stars undergo that 
particular, small probability event is far from clear. 

The major challenge that we believe this work reveals is that the MRI may have important effects
on the evolution of stars and that to truly appreciate its effect, one-dimensional ``shellular"
calculations of stellar evolution may not be adequate. The MRI, and other instabilities, are anisotropic
and non--axisymmetric. They are likely to be triggered in complex patterns in the star and to 
engender complex flow distributions. 

If magnetoratational effects are active in the later stages of stellar evolution, then the overall sign of the effect 
seems clear: the interior of stars will rotate more slowly, perhaps much more slowly, than rotating 
stellar evolution calculations in the absence of magnetic effects would indicate. Ironically, this might 
mean that legions of zero rotation or small rotation core-collapse calculations are more pertinent than 
one might have thought. 

\acknowledgments
We are grateful for discussions of the MRI and related issues with Steve Balbus, Henk Spruit, and Ethan 
Vishniac and to the referee, Kristin Menou, for valuable feedback that improved the manuscript.
We thank the MESA team for making this valuable tool readily available and especially thank Bill Paxton 
for his counsel in running the code. This work was begun at the Kavli Institute for Theoretical Physics that is
currently supported by NSF PHY11-25915. Some work on this paper was also done in the hospitable 
environment of the Aspen Center for Physics that is supported by NSF Grant PHY-1066293. JCW is 
especially grateful for the supportive staff and conducive environment of both KITP and the Aspen
Center for Physics. EC thanks the Enrico Fermi Institute for its support via the Enrico Fermi Postdoctoral 
Fellowship. This work was supported in part by NSF Grants AST-0707769 and NSF AST-1109801.

\newpage

\begin{center}
\begin{deluxetable}{lcccc}
\tablewidth{0pt}
\tablenum{1}
\tablecolumns{5}
\tablecaption{15 \ms\ Models\label{models}}
\tablehead{\colhead{Model} & \colhead{Final Mass} & \colhead{Final Fe Mass} & \colhead{Final Eq. Vel.} & \colhead{Final Fe Eq. Vel.} \\
                  \colhead{}          & \colhead{\ms}          &  \colhead{\ms}               & \colhead{\kms}            & \colhead{\kms} }
\startdata
No MRI, No ST & 12.7 & 1.29 & 0.07 & 620 \\
ST Only & 14.3 & 1.28 & 206 & $\sim 0$ \\
MRI Only & 13.2 & 1.22 & 0.12 & 270  \\
Both ST \& MRI & 14.3 & 1.34 & 0.10 & 21 \\
\enddata
\end{deluxetable}
\label{models}
\end{center}

\newpage

\newpage

\newpage

\newpage

\newpage

\newpage

\newpage

\newpage

\newpage

\newpage

\newpage

\newpage

\newpage

\newpage

\newpage

\newpage

\clearpage

\newpage

\newpage

\newpage

\end{document}